\documentclass[final,3p,times,sort&compress]{elsarticle}

\usepackage{graphicx}
\usepackage{amssymb}
\usepackage{graphicx}
\usepackage{dcolumn}
\usepackage{bm}
\usepackage{color}
\usepackage{graphicx}
\usepackage{multirow}
\usepackage{tabularx}
\usepackage{tikz}
\usepackage{pgfplots}
\usepackage{pgflibraryarrows}
\usepackage{pgflibrarysnakes}

\begin{document}

\begin{frontmatter}

\title{Early warning of large volatilities based on recurrence interval analysis in Chinese stock markets}

\author[ECUST,BU]{Zhi-Qiang Jiang}
\author[UFAL,BU]{Askery A. Canabarro}
\author[UR]{Boris Podobnik}
\author[BU]{H. Eugene Stanley}
\author[ECUST]{Wei-Xing Zhou\corref{cor1}}
\cortext[cor1]{Corresponding author. Address: 130 Meilong Road, P.O. Box 114, School of Business,
              East China University of Science and Technology, Shanghai 200237, China, Phone: +86 21 64250053, Fax: +86 21 64253152.}
\ead{wxzhou@ecust.edu.cn} %

\address[ECUST]{School of Business and Research Center for Econophysics, East China University of Science and Technology, Shanghai 200237, China}
\address[BU]{Department of Physics and Center for Polymer Studies, Boston University, Boston, MA 02215, USA}
\address[UFAL]{Grupo de F\'isica da Mat\'eria Condensada, N\'ucleo de Ci\^encias Exatas,
Campus Arapiraca, Universidade Federal de Alagoas, 57309-005, Arapiraca-AL, Brazil}
\address[UR]{Faculty of School Engineering, University of Rijeka, 51000 Rijeka, Croatia}

\begin{abstract}
Being able to forcast extreme volatility is a central issue in financial
risk management. We present a large volatility predicting method based
on the distribution of recurrence intervals between volatilities
exceeding a certain threshold $Q$ for a fixed expected recurrence time
$\tau_Q$. We find that the recurrence intervals are well approximated by
the $q$-exponential distribution for all stocks and all $\tau_Q$
values. Thus a analytical formula for determining the hazard probability
$W(\Delta t |t)$ that a volatility above $Q$ will occur within a short
interval $\Delta t$ if the last volatility exceeding $Q$ happened $t$
periods ago can be directly derived from the $q$-exponential
distribution, which is found to be in good agreement with the empirical
hazard probability from real stock data. Using these results, we adopt a
decision-making algorithm for triggering the alarm of the occurrence of
the next volatility above $Q$ based on the hazard probability. Using a
``receiver operator characteristic'' (ROC) analysis, we find that this
predicting method efficiently forecasts the occurrance of large
volatility events in real stock data. Our analysis may help us better
understand reoccurring large volatilities and more accurately quantify
financial risks in stock markets.
\medskip
\noindent {\textit{JEL classification}}: C14
\end{abstract}

\begin{keyword}
Extreme volatility, Risk estimation, Recurrence interval; Large
volatility forecasting; Distribution; Hazard probability
\end{keyword}

\end{frontmatter}



\section{Introduction}
\label{S1:Introduction}

Predicting extreme volatility events in financial markets is essential
when estimating risk. A standard approach to extreme event prediction is
to find the precursory patterns prior to an extreme event or to quantify
the probability that a given pattern is a precursor to an extreme event
\citep{Hallerberg-Altmann-Holstein-Kantz-2007-PRE,Hallerberg-Kantz-2008-PRE}.
\cite{Bogachev-Bunde-2009-PRE,Bogachev-Bunde-2011-PA} propose a new method based on the statistics of the recurrence
intervals between events exceeding a threshold to determine the risk
probability $W(\Delta t | t)$ that an extreme event will occur within
the next $\Delta t$ intervals when the last extreme event occurred $t$
periods ago. They
find that when examining real market data and model data with a low
level of noise the predicting method based on recurrence interval
analysis produces a better performance forecast than the method based on
precursor pattern recognition
\citep{Bogachev-Bunde-2009-PRE,Bogachev-Bunde-2011-PA}.

Understanding of the recurrence interval, defined as the waiting time
between consecutive events with values greater than a predefined
threshold $Q$, is essential in uncovering the underlying laws governing
extreme events in many fields. Recurrent interval analysis has been
carried out on many kinds of time series in predicting the probability
that an extreme event will occur, including records of climate
\citep{Bunde-Eichner-Havlin-Kantelhardt-2004-PA,Bunde-Eichner-Kantelhardt-Havlin-2005-PRL},
seismic activities \citep{Saichev-Sornette-2006-PRL}, energy dissipation
rates of three-dimensional turbulence
\citep{Liu-Jiang-Ren-Zhou-2009-PRE}, heartbeat intervals in medical
science \citep{Bogachev-Kireenkov-Nifontov-Bunde-2009-NJP},
precipitation and river runoff \citep{Bogachev-Bunde-2012-EPL}, internet
traffic \citep{Bogachev-Bunde-2009-EPL,Cai-Fu-Zhou-Gu-Zhou-2009-EPL},
financial volatilities
\citep{Yamasaki-Muchnik-Havlin-Bunde-Stanley-2005-PNAS,Xie-Jiang-Zhou-2014-EM},
equity returns
\citep{Yamasaki-Muchnik-Havlin-Bunde-Stanley-2006-inPFE,Bogachev-Eichner-Bunde-2007-PRL,Bogachev-Bunde-2008-PRE,Bogachev-Bunde-2009-PRE,Ren-Zhou-2010-NJP,Ludescher-Tsallis-Bunde-2011-EPL,He-Chen-2011b-PA,Meng-Ren-Gu-Xiong-Zhang-Zhou-Zhang-2012-EPL,Suo-Wang-Li-2005-EM},
and trading volumes
\citep{Podobnik-Horvatic-Petersen-Stanley-2009-PNAS,Ren-Zhou-2010-PRE,Li-Wang-Havlin-Stanley-2011-PRE}.
An improved method of estimating value at risk (VaR) in financial
markets has been proposed based on the recurrence interval between the
last two returns below $-Q$. This method is significantly more accurate
than traditional estimates based on the overall or local return
distributions
\citep{Bogachev-Bunde-2009-PRE,Ludescher-Tsallis-Bunde-2011-EPL}.

To accurately estimate the risk probability and the VaR based on
recurrence interval analysis we need the set of distribution and
memory behavior of the recurrence time between extreme events. It is found that
the recurrence intervals of the time series in many different fields exhibit fat-tailed distributions and long- and short-range memories,
indicating that extreme events are not described by the Poisson
process. Unlike long- and short-range memory behaviors, which are easily
testable using conditional distribution analysis and the DFA method, the
distribution form of recurrence intervals is still elusive.  For
example, in financial markets the recurrence intervals of different data
types (return, volatility, and trading volume), different data
resolutions (minute-by-minute and daily), and different markets fit
different distributions, including power-law, stretched exponential, and
$q$-exponential.  It has been found that the recurrence interval
distribution above a fixed threshold has a power-law tail for the daily
volatilities in Japanese market
\citep{Kaizoji-Kaizoji-2004a-PA,Yamasaki-Muchnik-Havlin-Bunde-Stanley-2005-PNAS},
the minute-by-minute volatilities in Korean market
\citep{Lee-Lee-Rikvold-2006-JKPS} and Italian market
\citep{Greco-SorrisoValvo-Carbone-Cidone-2008-PA}, the daily returns in
US stock markets
\citep{Bogachev-Eichner-Bunde-2007-PRL,Bogachev-Bunde-2008-PRE,Bogachev-Bunde-2009-PRE},
the minute-by-minute returns in Chinese markets
\citep{Ren-Zhou-2010-NJP}, and the minute-by-minute trading volume in US
markets \citep{Li-Wang-Havlin-Stanley-2011-PRE} and Chinese markets
\citep{Ren-Zhou-2010-PRE}. A number of studies ranging from daily to
high-frequency data and from developed to emerging markets
\citep{Wang-Yamasaki-Havlin-Stanley-2006-PRE,Wang-Weber-Yamasaki-Havlin-Stanley-2007-EPJB,Jung-Wang-Havlin-Kaizoji-Moon-Stanley-2008-EPJB,Qiu-Guo-Chen-2008-PA,Ren-Gu-Zhou-2009-PA,Ren-Guo-Zhou-2009-PA,Jeon-Moon-Oh-Yang-Jung-2010-JKPS,Wang-Wang-2012-CIE,Xie-Jiang-Zhou-2014-EM},
have also reported that the distribution of the recurrence intervals of
financial volatility is a stretched
exponential. Reference~\cite{Suo-Wang-Li-2005-EM} reports that in
Chinese markets the recurrence time between returns above a given
positive threshold or below a negative threshold for the index spot and
futures fits a stretched exponential distribution.  The recurrence
intervals between losses in financial markets were recently found to fit
a $q$-exponential distribution
\citep{Ludescher-Tsallis-Bunde-2011-EPL,Ludescher-Bunde-2014-PRE}.

In this paper we describe the datasets in Sec.~\ref{S1:Data}, present
the theoretical framework for predicting large volatilities in
Sec.~\ref{S1:Framework}, determine the distribution of the recurrence
intervals between large volatilities in Sec.~\ref{S1:Distribution}, and
report the hazard probability results and predicting algorithm
performance in Sec.~\ref{S1:Risk}.  In Sec.~\ref{S1:Conclusion} we
summarize our findings.

\section{Data description}
\label{S1:Data}

To carry out a detailed recurrence interval analysis of Chinese stock
markets, we include as many Chinese stocks in our analyzing sample as
possible. The minute-by-minute price data of all stocks in the Chinese
markets are extracted from the RESSET financial database. The extracting
period is from 26 July 1999 to 30 December 2011, which is the maximum
spanning period allowed in the RESSET database. To ensure that the
recurrent interval results between the top 1\% volatilities will have
more than 1,000 data points, we select only those stocks that have a
minimum of two years of trading records. Having this large a sample size
lowers the error rate when we use a maximum likelihood estimation to fit
the distributions.  Finally, we have 1891 stocks in our sample, which include 853 A-shares,
54 B-shares, and 63 ChiNext shares in the Shenzhen market, and 867
A-shares and 54 B-shares in the Shanghai market.

\section{Framework of predicting large volatilities}
\label{S1:Framework}
\subsection{Hazard probability $W(\Delta t | t)$}

We use the hazard probability $W(\Delta t | t)$ to forecast the
occurrance of large volatility events. The $W(\Delta t | t)$ is the
probability that there will be additional waiting time $\Delta t$ before
another large volatility event occurs when the previous large volatility
event occurred $t$ time ago. This probability is the key early-warning
measurement for the occurrence of extreme volatilities. The early
warning is triggered when the probability $W(\Delta t | t)$ is greater
than a predefined alarm threshold. We can theoretically derive this
hazard probability if we have the distribution of the time intervals
between consecutive extreme volatilities, which are defined as the
volatilities that exceed a given threshold $Q$.

Using the probability density $p(t)$ of the recurrence intervals between
the extreme volatilities, if the time elapsed since the last extreme
event is $t$, we want to determine the probability density function
$p(\Delta{t}|t)$ that quantifies the additional waiting time $\Delta{t}$
until the next extreme event. Using the Bayes theorem for conditional
probabilities, the probability that an event A occurs, given the
knowledge of an event B, is simply the quotient of the probability of
the event A without constraint and the probability of event B
\citep{Sornette-Knopoff-1997-BSSA},
\begin{equation}
  p(A|B) = \frac{p(AB)}{p(B)},
  \label{Eq:P:A:B}
\end{equation}
where $p(AB)=p(t+\Delta{t})$ is the probability that no event occurs
from 0 to $t$ and that an event occurs at $t+\Delta{t}$, and
$p(B)=\int_t^{\infty}p(s)ds$ is the probability that no event occurs
from 0 to $t$. Thus we have
\begin{equation}
  p(\Delta{t}|t) = \frac{p(t+\Delta{t})}{\int_t^{\infty}p(s)ds}.
  \label{Eq:P:Dt:t}
\end{equation}
Thus the hazard probability $W(\Delta{t},t)$ that an extreme event will
occur after a short time $\Delta{t}\ll t$ since the occurrance of the
previous extreme event can be expressed
\begin{equation}
 W(\Delta{t}|t)=\frac{\int_t^{t+\Delta{t}}p(t)dt}{\int_t^{\infty}p(t)dt}.
 \label{Eq:Wq}
\end{equation}

\subsection{Predicting algorithm}

For a given distribution $p(t)$ of the recurrence intervals between
extreme volatilities, the formula for $W(\Delta{t}|t)$ can be obtained
using equation~(\ref{Eq:Wq}). If $\Delta t=1$ for $W(\Delta{t}|t)$, the
hazard probability and a decision-making algorithm can be used to
predict large volatilities \citep{Bogachev-Bunde-2011-PA}. To trigger an
early warning that a large volatility is about to occur, we set a
threshold $Q_p$ for the hazard probability. When the hazard probability
exceeds $Q_p$, an alarm that a large volatility will occur during the
next time point is activated. We next estimate the $Q_p$ parameter,
which is the maximum correct prediction rate when the maximum
false-alarm tolerance is set.

To determine $Q_p$ we estimate the correct prediction and false alarm
rates for each $Q_p$ in the range of $[0, 1]$. We then plot the correct
prediction rate with respect to the false alarm rate and get the
``receiver operator characteristic'' (ROC) curve
\citep{Bogachev-Bunde-2009-PRE,Bogachev-Bunde-2009-EPL,Bogachev-Kireenkov-Nifontov-Bunde-2009-NJP,Bogachev-Bunde-2011-PA}.
The ROC curve is used to quantify prediction efficiency. The satisfied
$Q_p$ corresponds to the point at which the false alarm rate equals the
tolerant alarm level on the ROC curve.

To estimate the correct prediction and false alarm rates we generate for
a given $Q_p$ two forecasting signals---alarms and non-alarms---at each
time point.  By comparing the forecasting signals with the real data, we
obtain one of four outcomes at each time point
\citep{Bogachev-Bunde-2011-PA}, (i) a correct prediction of a large
volatility event, (ii) a correct prediction of a non-large volatility
event, (iii) a missed event, and (iv) a false alarm. By recording in our
testing records how many times each outcome occurs we can estimate the
correct prediction rate $D$ and the false alarm rate $A$ using
\begin{equation}
 D = \frac{O_{11}}{ O_{01} + O_{11}},~~A = \frac{O_{10}}{O_{00} + O_{10}},
 \label{Eq:Wq:AD}
\end{equation}
where $O_{11}$ is the number of large volatility events that are
correctly predicted, $O_{00}$ the number of non-large volatility events
that are correctly predicted, $O_{01}$ the number of missed events, and
$O_{10}$ the number of false alarms. All possible pairs of $(D, A)$ will
be obtained if we vary the $Q_p$ range from 0 to 1.

By definition the ROC curve will be $D=A=1$ if $Q_p =0$ and $D=A=0$ if
$Q_p = 1$. Note that the ROC curve joins the point $(0, 0)$ in the left
bottom corner to the point $(1, 1)$ in the right top corner. Note also
that, for the random guess outcome, $D=A$, a straight line between the two
corners. This occurs when there is no memory in the data. For a fixed
value of $A$, the larger the value of the correct prediction rate, the
better this algorithm performs.

\section{Distribution of recurrence intervals between large volatilities}
\label{S1:Distribution}

\subsection{Definition of volatilities and recurrence intervals}

For a given minute-by-minute price series $p(t)$, the minute-by-minute
volatility $\omega(t)$ can be estimated \citep{Bollen-Inder-2002-JEF}
using
\begin{equation}
  \omega(t)=|\ln p(t)-\ln p(t-1)|.
  \label{vlt}
\end{equation}
In order to eliminate the influence of the daily periodic patterns, we
remove the intraday patterns from the volatility series $\omega(t)$ on
each trading day,
\begin{equation}
  \omega'(s)=\omega(s)/A(s),
  \label{vltrip}
\end{equation}
where $A(s) = \sum_i^N \omega(i, s) / N$. Here $\omega(i, s)$ represents
the volatility at time $s$ on day $i$. The normalized volatility series
$v(t)$ is then obtained by dividing $\omega'(t)$ by its standard
deviation,
\begin{equation}
  v(t)= \frac{\omega'(t)}{ \sqrt{[ \langle \omega'(t)^2 \rangle - \langle \omega'(t) \rangle^2 ]} }.
  \label{nvlt}
\end{equation}

The focus of our study is the recurrence interval between the normalized
volatilities exceeding a predefined threshold $Q$. To compare the
results between different stocks, we quantify $Q$ by its mean recurrence
time $\tau_Q$. There is a one-to-one correspondence between $Q$ and
$\tau_Q$, such that \citep{Podobnik-Horvatic-Petersen-Stanley-2009-PNAS}
\begin{equation}
  \frac{1}{\tau_Q} = \int_Q^\infty p(v) {\rm{d}}v,
  \label{tauQpv}
\end{equation}

where $p(v)$ is the probability distribution of the volatility. Here we
restrict $\tau_Q$ to a range of $[20, 100]$. This range corresponds to
the extreme volatilities from a top value of \%5 to 1\%, which is often
considered in the risk estimation.

\subsection{Distribution formula of recurrence intervals}

\begin{figure}[htb]
\centering
\includegraphics[width=8cm]{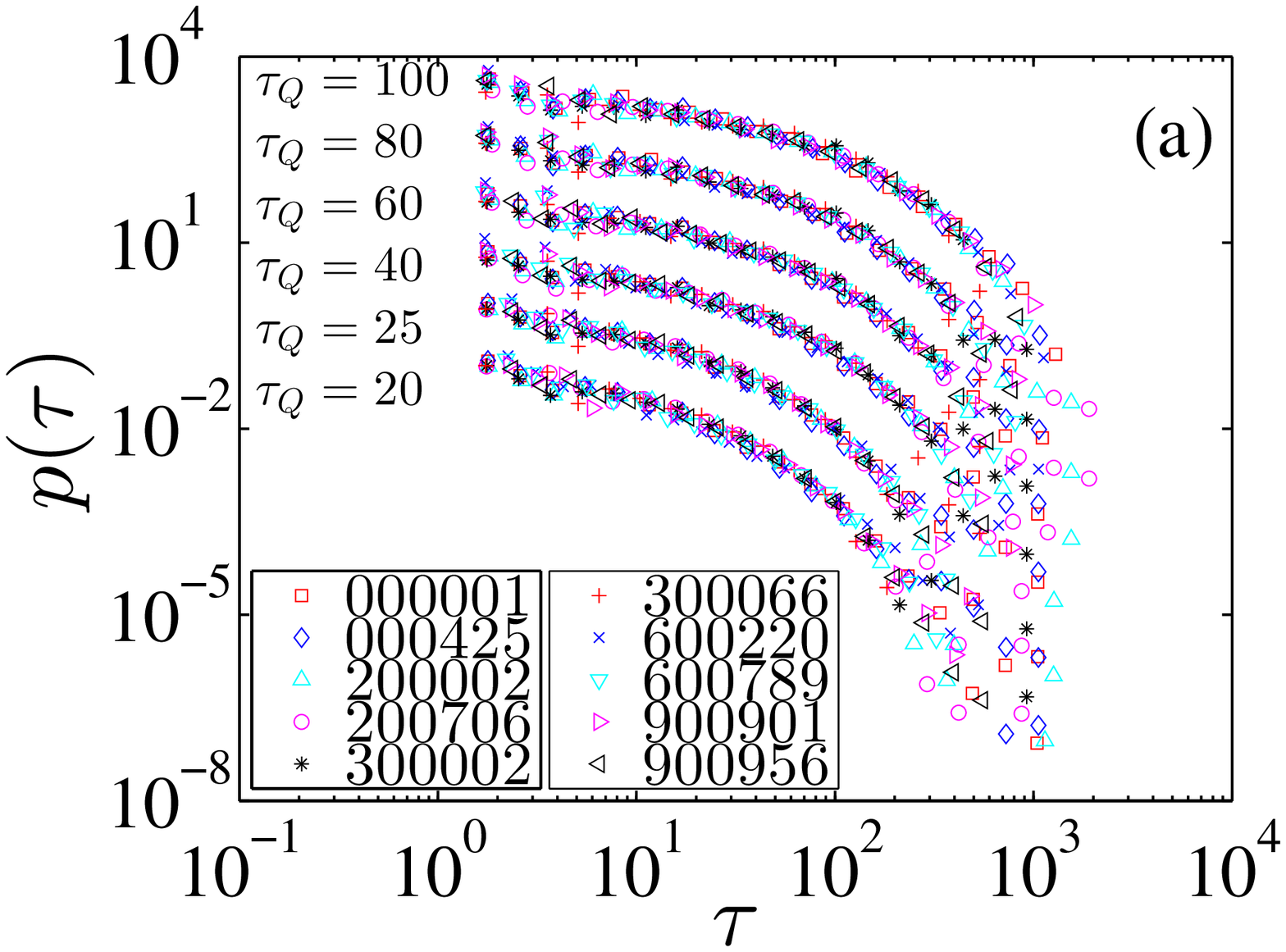}
\includegraphics[width=7.3cm]{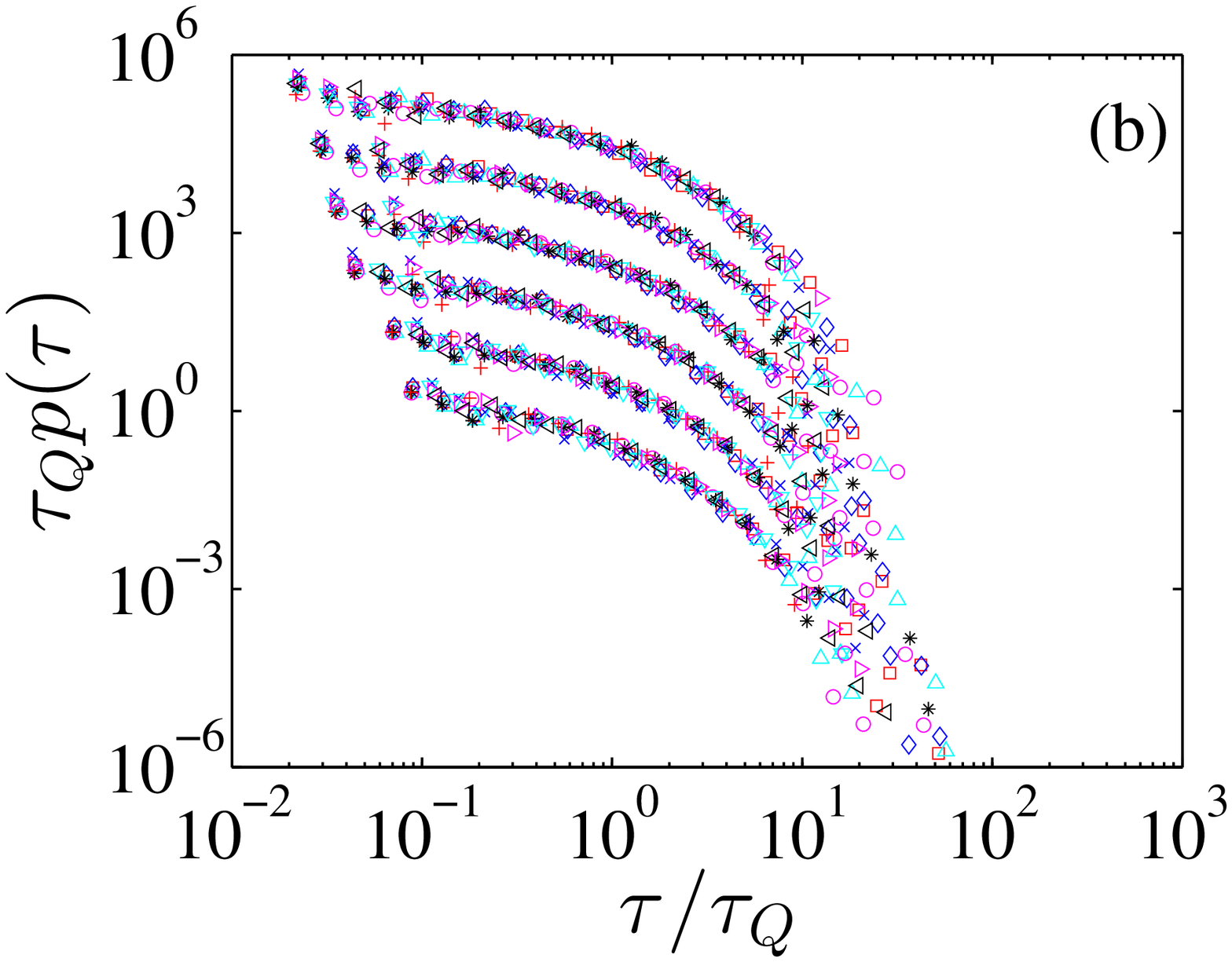}
\caption{\label{Fig:RI:PDF} (color online). Probability distribution of
  the recurrence intervals for 10 randomly chosen stocks. For better
  visibility, the distribution curve of $\tau_Q =25$, $40$, $60$, $80$,
  and $100$ are shifted vertically by a factor of 10, 100, 1000, 10000,
  and 100000, respectively.  (a) Original recurrence intervals. (b)
  Scaled recurrence intervals.}
\end{figure}

To analytically determine the hazard probability we find the
distribution that best approximates the recurrence interval distribution
for all the stocks in our sample. We also determine whether the
distribution parameters are dependent on the mean recurrence interval
$\tau_Q$ and on whether the market is bear or bull. Previous research
has indicated that the distribution of recurrence intervals between
returns below a negative threshold $-Q$ depends only on the mean
recurrence interval $\tau_Q$, and not on a specific asset or on the time
resolution of the data
\citep{Ludescher-Tsallis-Bunde-2011-EPL,Ludescher-Bunde-2014-PRE}. We
first check whether the recurrence time between volatilities above a
threshold $Q$ exhibits this behavior. Figure~\ref{Fig:RI:PDF}(a) shows
the probability distribution of the recurrence intervals for 10 randomly
chosen stocks. We transform the distribution curves of different
$\tau_Q$ values by a factor to increase visibility. Note that for the
same $\tau_Q$ value the recurrence time distributions of different
stocks nearly overlap on the same curve. This means that the return
intervals between volatilities that exceed a threshold $Q$ may exhibit a
universal distribution for different stocks when $\tau_Q$ is fixed. We
also want to know whether the distributions of different $\tau_Q$ values
share the same pattern. Previous research indicates that these
distributions are influenced only by the mean recurrence time $\tau_Q$
for return recurrence intervals
\citep{Ludescher-Bunde-2014-PRE}. Although the distributions in
Fig.~\ref{Fig:RI:PDF}(a) seem to be different for different $\tau_Q$, if
we scale the distribution using the mean recurrence time $\tau_Q$ the
six distributions seem to be parallel [see Fig.~\ref{Fig:RI:PDF}(b)].

To quantitatively measure how the recurrence interval
distributions vary with the mean interval $\tau_Q$, we need a
suitable distributional formula for capturing the recurrence interval
distribution. Previous research shows that the recurrence intervals can be fitted
by the stretched distribution
\citep{Wang-Yamasaki-Havlin-Stanley-2006-PRE,Wang-Weber-Yamasaki-Havlin-Stanley-2007-EPJB,Jung-Wang-Havlin-Kaizoji-Moon-Stanley-2008-EPJB,Qiu-Guo-Chen-2008-PA,Jeon-Moon-Oh-Yang-Jung-2010-JKPS,Wang-Wang-2012-CIE,Suo-Wang-Li-2005-EM},
the power-law distribution with an exponential cutoff
\citep{Kaizoji-Kaizoji-2004a-PA,Yamasaki-Muchnik-Havlin-Bunde-Stanley-2005-PNAS,Lee-Lee-Rikvold-2006-JKPS,Greco-SorrisoValvo-Carbone-Cidone-2008-PA},
and the $q$-exponential distribution
\citep{Ludescher-Tsallis-Bunde-2011-EPL,Ludescher-Bunde-2014-PRE}.
The Weibull distribution is usually comparable to the $q$-exponential distribution
\citep{Poloti-Scalas-2008-PA,Jiang-Chen-Zhou-2008-PA}, and we use both
two-parameter and three-parameter Weibull distributions to fit
the recurrence intervals in our analysis. The following are five
candidate distributions: the stretched exponential distribution,
\begin{equation}
 p(\tau) = a \exp[-(b \tau)^\mu],
 \label{Eq:StrExp:Tau:PDF}
\end{equation}
the power-law distribution with an exponential cutoff,
\begin{equation}
 p(\tau) = c \tau^{-\gamma-1} \exp(-k \tau),
 \label{Eq:PowExp:Tau:PDF}
\end{equation}
the $q$-exponential distribution,
\begin{equation}
 p(\tau) = (2-q) \lambda [1+(q-1) \lambda \tau]^{-\frac{1}{q-1}},
 \label{Eq:qExp:Tau:PDF}
\end{equation}
the two-parameter Weibull distribution,
\begin{equation}
 p(\tau) = \frac{\zeta}{d} \left( \frac{\tau}{d}\right)^{\zeta-1}\exp
 \left[-\left( \frac{\tau}{d}
\right)^\zeta \right],
 \label{Eq:2WBL:Tau:PDF}
\end{equation}
and the three-parameter Weibull distribution,
\begin{equation}
 p(\tau) =\frac{\zeta}{d} \left( \frac{\tau -
   \tau_0}{d}\right)^{\zeta-1}\exp \left[- \left( \frac{\tau -
     \tau_0}{d}\right)^\zeta  \right].
 \label{Eq:3WBL:Tau:PDF}
\end{equation}

To compare the recurrence interval distributions of different $\tau_Q$
values, we normalize the recurrence intervals by $\tau_Q$, i.e., $x =
\tau / \tau_Q$. We obtain the five candidate distributions used to fit
the normalized recurrence intervals by substituting $\tau = x \tau_Q$
into $f(x) = p(\tau) \tau_Q$, i.e.,
\begin{equation}
f(x) = a \bar{\tau} \exp[-(b \tau_Q x)^\mu],
 \label{Eq:StrExp:x:PDF}
\end{equation}
\begin{equation}
f(x) = c \tau_Q^{-\gamma} x^{-\gamma-1} \exp(-k \tau_Q x),
 \label{Eq:PowExp:x:pdf}
\end{equation}
\begin{equation}
 p(\tau) = (2-q) (\lambda \tau_Q) [1+(q-1) (\lambda \tau_Q)
   \tau]^{-\frac{1}{q-1}},
 \label{Eq:qExp:x:PDF}
\end{equation}
\begin{equation}
 f(x) = \frac{\zeta}{d/\tau_Q} \left(
 \frac{x}{d/\tau_Q}\right)^{\zeta-1}\exp\left[-\left(
   \frac{x}{d/\tau_Q}\right)^\zeta \right],
 \label{Eq:2WBL:x:pdf}
\end{equation}
and
\begin{equation}
f(x) =\frac{\zeta}{d/\tau_Q} \left( \frac{x -
  \tau_0/\tau_Q}{d/\tau_Q}\right)^{\zeta-1}\exp\left[-\left( \frac{x -
    \tau_0/\tau_Q}{d/\tau_Q}\right)^\zeta \right].
 \label{Eq:3WBL:x:pdf}
\end{equation}

We use the maximum likelihood estimation (MLE) method to estimate the
parameters of the five distribution parameters. The details of fitting
the stretched exponential distribution and the power-law distribution
with an exponential cutoff are presented in the Appendix. Note that in
the following analysis we fit only the scaled recurrence intervals.

\begin{figure}[htb]
\centering
\includegraphics[width=8cm]{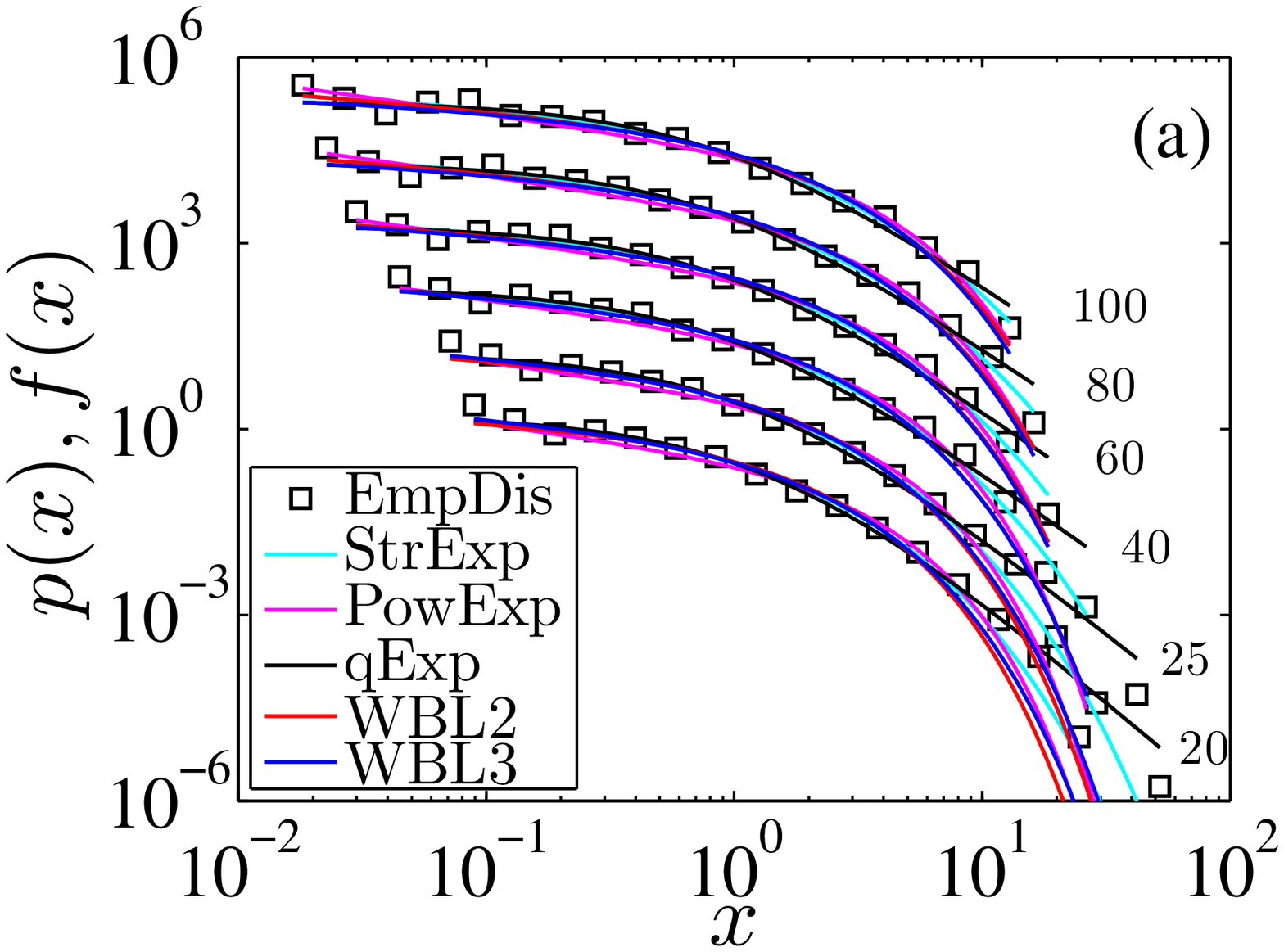}
\includegraphics[width=8cm]{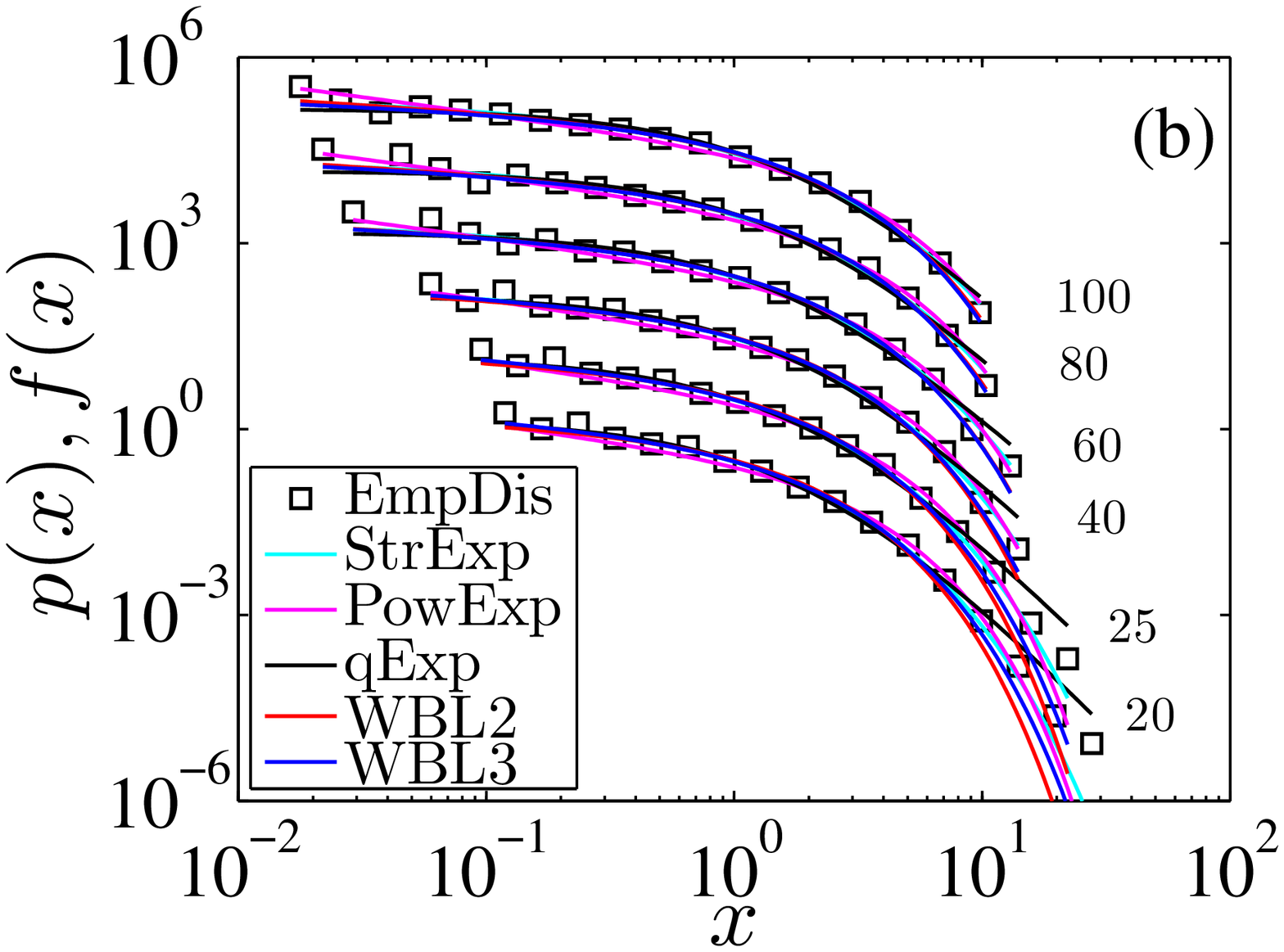}
\caption{\label{Fig:RI:PDF:Fit} (color online). Plots of distribution
  fits to normalized recurrence intervals for two stocks. The open
  markers are the empirical distributions and the solid lines are the
  fits to the five candidate distributions. For better visibility, the
  curves of $\tau_Q =25$, $40$, $60$, $80$, and $100$ are shifted
  vertically by a factor of 10, 100, 1000, 10000, and 100000,
  respectively.  (a) Stock 000001. (b) Stock 900956.}
\end{figure}

\begin{figure}[htb]
\centering
\includegraphics[width=5cm]{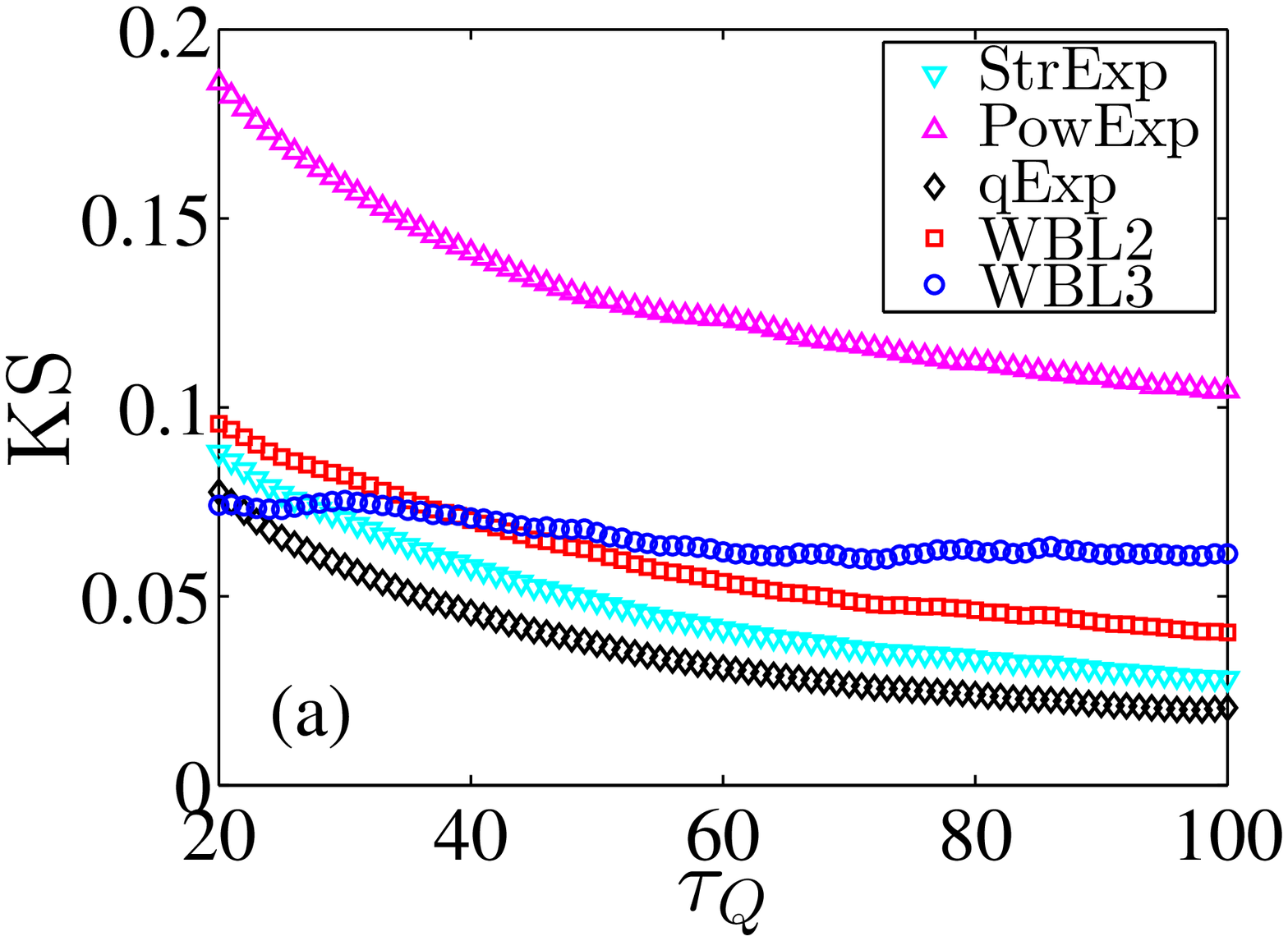}
\includegraphics[width=5cm]{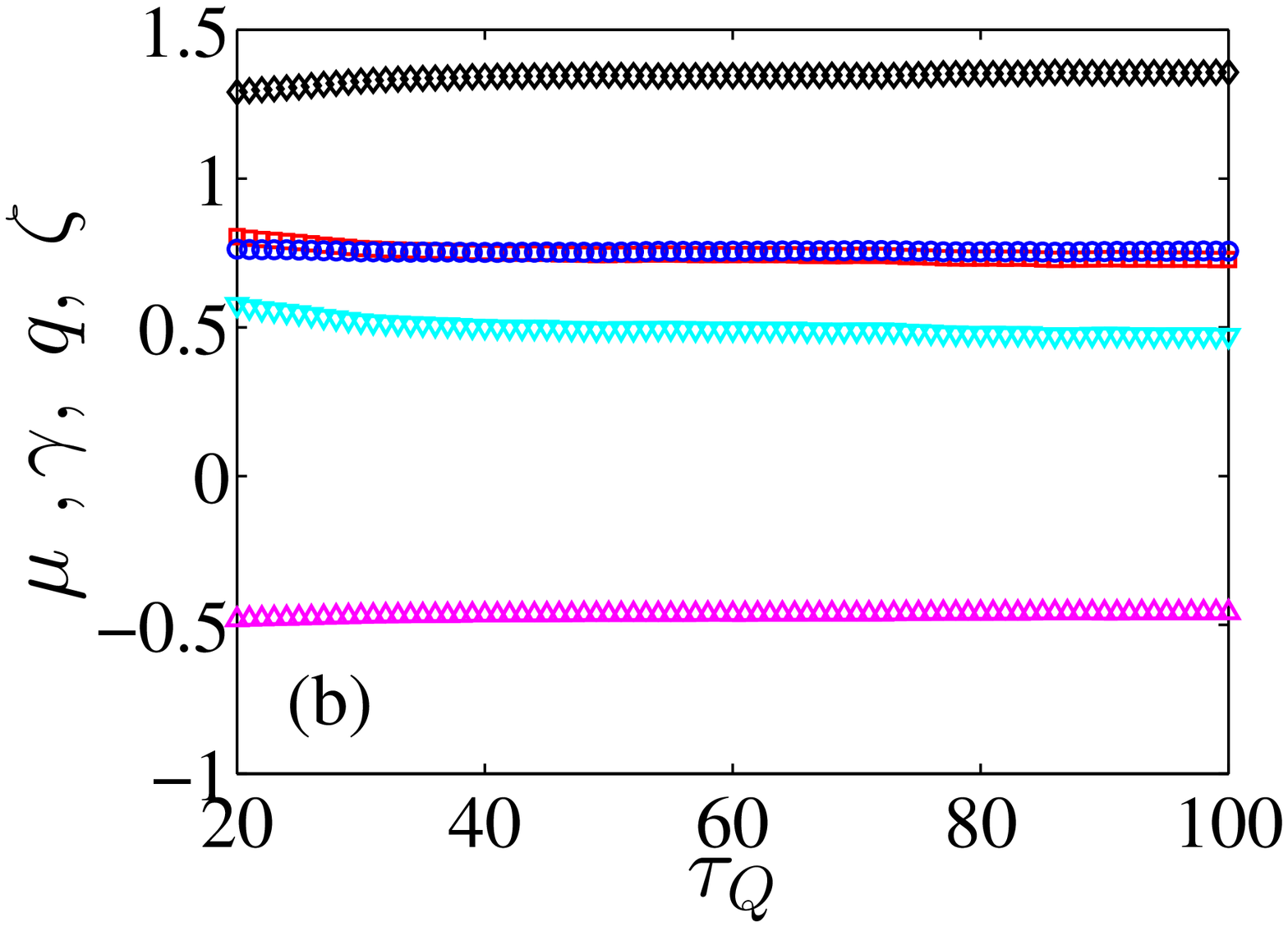}
\includegraphics[width=5cm]{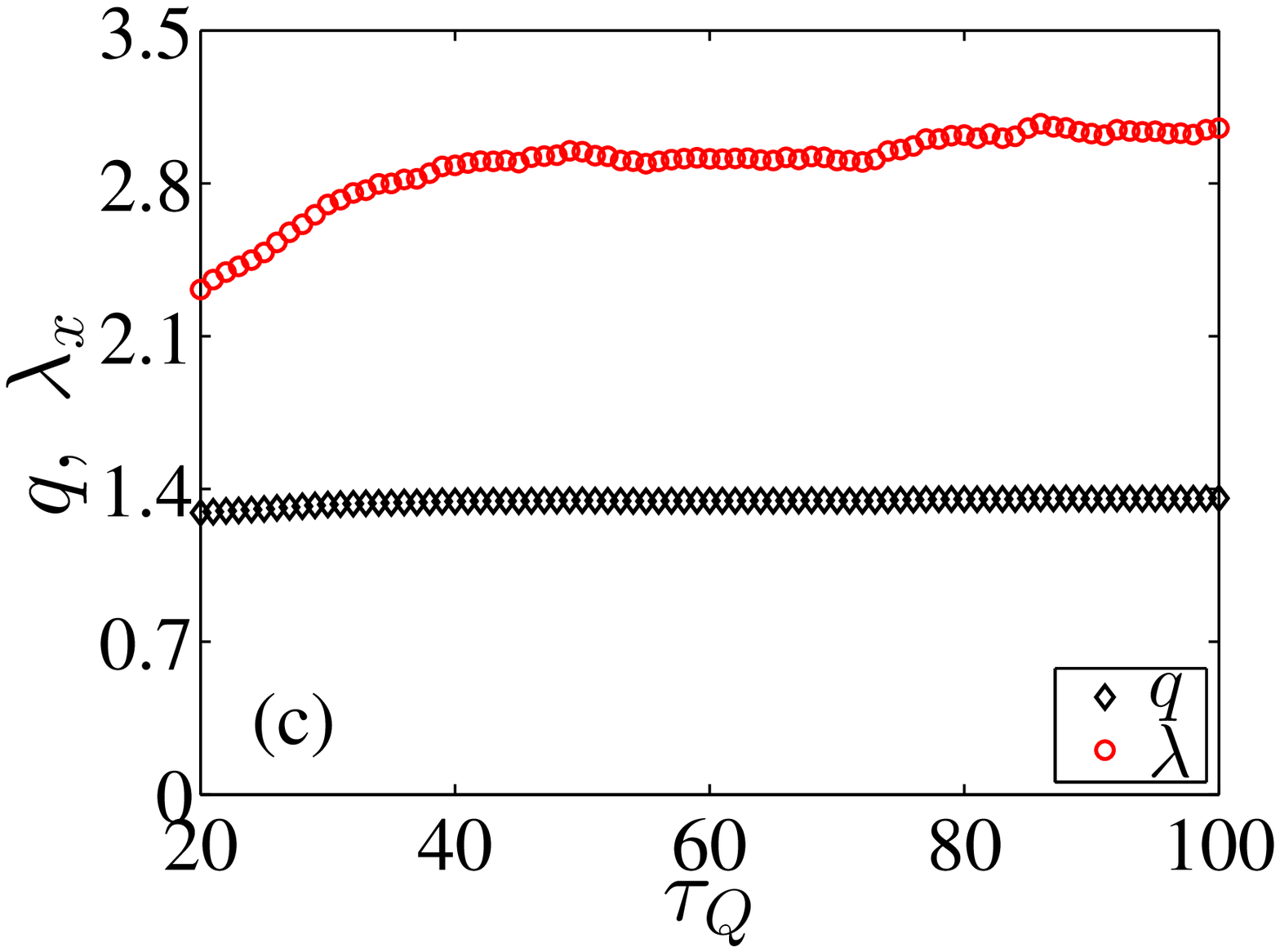}
\includegraphics[width=5cm]{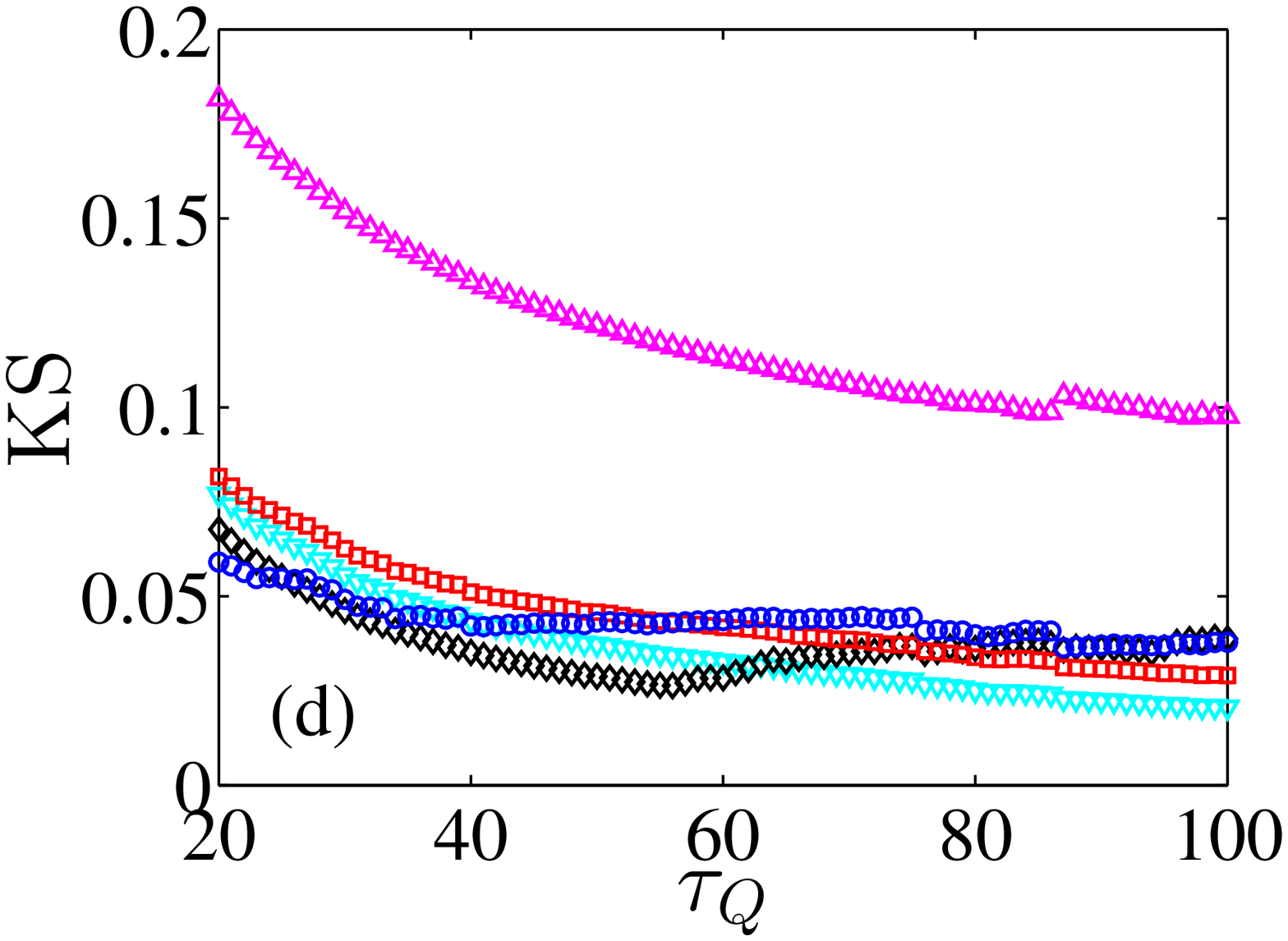}
\includegraphics[width=5cm]{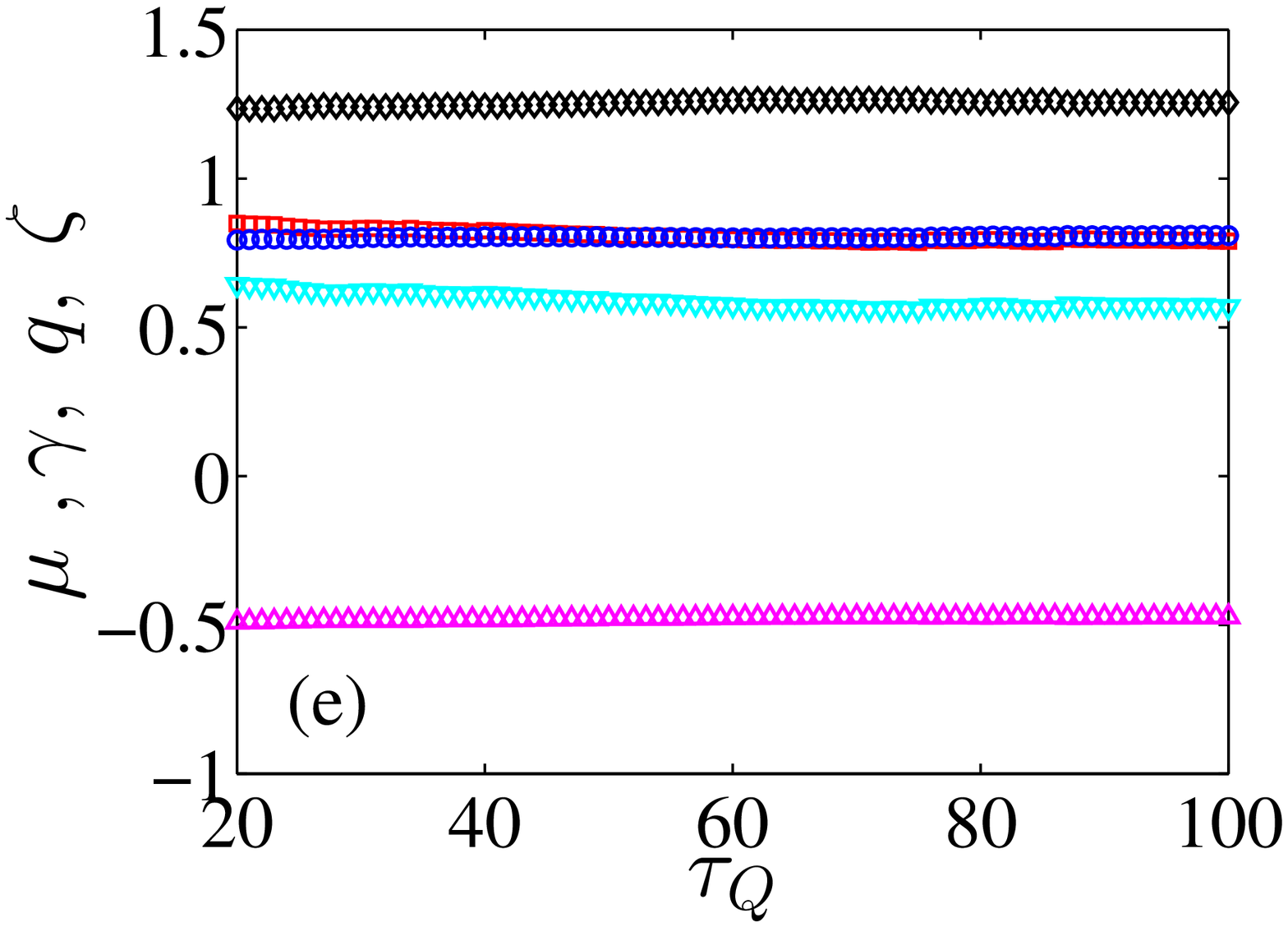}
\includegraphics[width=5cm]{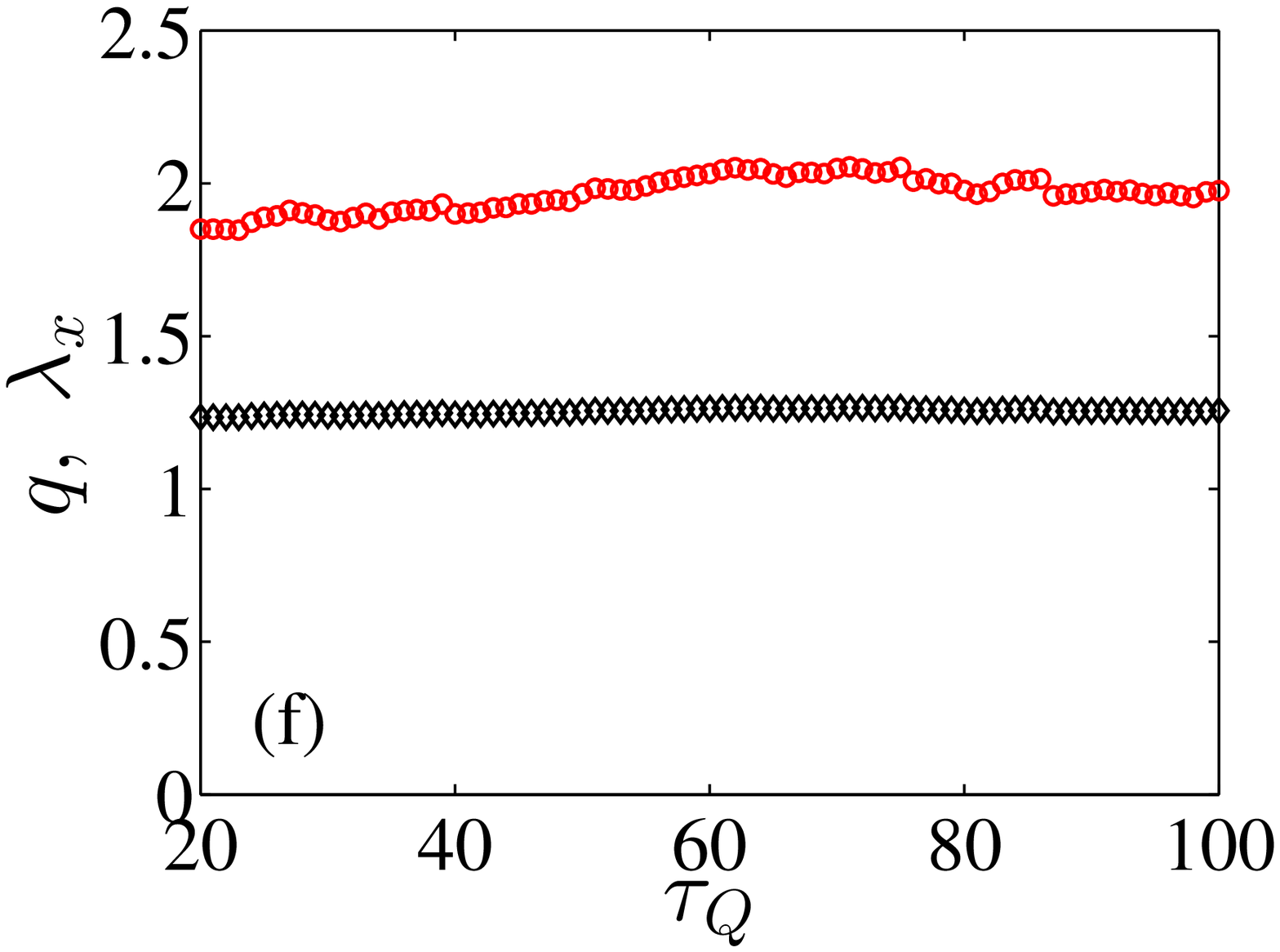}
\caption{\label{Fig:RI:Fit:Par} (color online). Plots of fitting results
  for two stocks. (a-c) Stock 000001. (d-f) Stock 900956. (a, d) Plots
  of the KS statistics with respect to the mean interval $\tau_Q$. The
  KS statistics is used to describe the agreement between the empirical
  distribution and the fitting distributions. (b, e) Comparison of the
  characteristic fitting parameters for the five candidate
  distributions. (c, f) Plots of the fitting parameters $\lambda_x =
  \tau_Q\lambda$ and $q$ with respect to $\tau_Q$. }
\end{figure}

Figure~\ref{Fig:RI:PDF:Fit} shows the empirical distributions and the
fitting results of the five candidate distributions of the scaled
recurrence intervals for two stocks, 000001 and 900956. Note that in the
central regions of the distributions all five candidates agree with the
empirical data. Note also that the $q$-exponential distribution better
fits the distribution tail for all $\tau_Q$ than any other candidate
distribution.  To determine which distribution has the best performance,
we utilize KS statistics to quantify the agreement between the empirical
distribution and the fitting distributions.
Figures~\ref{Fig:RI:Fit:Par}(a) and \ref{Fig:RI:Fit:Par}(d) show the KS
statistics of the five candidate distributions with respect to the mean
recurrence time $\tau_Q$ for the two stocks. For stock 000001 the
$q$-exponential distribution outperforms the other distributions and
possesses the smallest KS statistics for all $\tau_Q$. For stock 900956
the $q$-exponential distribution is best when $\tau_Q \le 60$, and the
stretched exponential distribution is best when $\tau_Q >
60$. Figure~\ref{Fig:RI:Fit:Par} shows plots of the characteristic
parameters of the five candidate distributions with respect to the mean
interval $\tau_Q$ for (b) stock 000001 and (e) stock 900956. Note that
all the fitting parameters of the five candidate distributions are
independent of the mean recurrence time $\tau_Q$ and exhibit a
horizontal line. This indicates that the distributions of the recurrence
intervals are not influenced by the threshold $Q$ when $\tau_Q$ is in
the $[20, 100]$ range. Figures~\ref{Fig:RI:Fit:Par}(c) and
\ref{Fig:RI:Fit:Par}(f) show a plot of the fitting parameters
$\lambda_x$, defined as $\tau_Q \lambda$, and $q$ as a function of
$\tau_Q$. Note that the fluctuations of $\lambda_x$ are not wide when
$\tau_Q > 40$ for stock 000001 and in the whole range of $\tau_Q$ for
stock 900956. These results indicate a scaling behavior in the
volatility recurrence intervals under different thresholds.

Our goal is to find a distribution that can approximate the distribution
of recurrence time. The best candidate distribution will be the one
that provides accuracy of fit and ease in estimating the hazard function
$W(t|\Delta t)$. Note that the power-law distribution with an
exponential cutoff provides the largest body of KS statistics for all
the stocks, and that the three-parameter Weibull distribution has fewer
KS statistics but requires that too many parameters be estimated. These
two distributions are excluded from the candidate list.  Because the
$q$-exponential distribution has the smallest body of KS statistics for
75.7\% of the fits, we use it to capture the distribution of the
recurrence intervals.  Even when the fits are of bodies of KS statistics
that are not the smallest, the $q$-exponential still provides a good
approximation of the recurrence time distribution. The $q$-exponential
distribution is also highly useful because it allows the derivation of
the analytical formula of the hazard function $W(t|\Delta t)$ from the
$q$-exponential formula.

\subsection{More on distribution parameter behaviors}

For each stock and each $\tau_Q$ value we fit the corresponding scaled
recurrence time $x$ with the $q$-exponential distribution and estimate
the distribution parameters $q$ and $\lambda_x$. To quantitatively check
the scaling behaviors we verify whether the parameters $q$ and
$\lambda_x$ are independent of $\tau_Q$ for the same stock and whether
they are the same across different stocks for the same $\tau_Q$. For
each stock we linearly regress the fitting parameters $q$ and
$\lambda_x$ with respect to $\tau_Q$. Although for $q$ vs $\tau_Q$ we
find that the maximum absolute slope is 0.006 and the mean slope over
all stocks $(5.68 \pm 7.77)\times 10^{-4}$, we do not observe comparable
small slopes for all stocks when fitting $\lambda_x$ vs $\tau_Q$. We
find 945 stocks with absolute slopes $<0.006$, which is the maximum
absolute slope of $q$ vs $\tau_Q$. The maximum value of the slope over
all stocks is 0.439 and the mean slope $0.0068 \pm 0.0213$ for
$\lambda_x$ vs $\tau_Q$. These results suggest that the parameter $q$ is
independent of the mean recurrence time $\tau_Q$ for all stocks and the
parameter $\lambda_x$ is independent of $\tau_q$ for half of the stocks,
which suggests that the scaling behavior in the recurrence intervals for
different $\tau_Q$ values exists in half of the stocks in the Chinese
markets ($945/1891 \approx 50\%$).

\begin{figure}[htb]
\centering
\includegraphics[width=5cm]{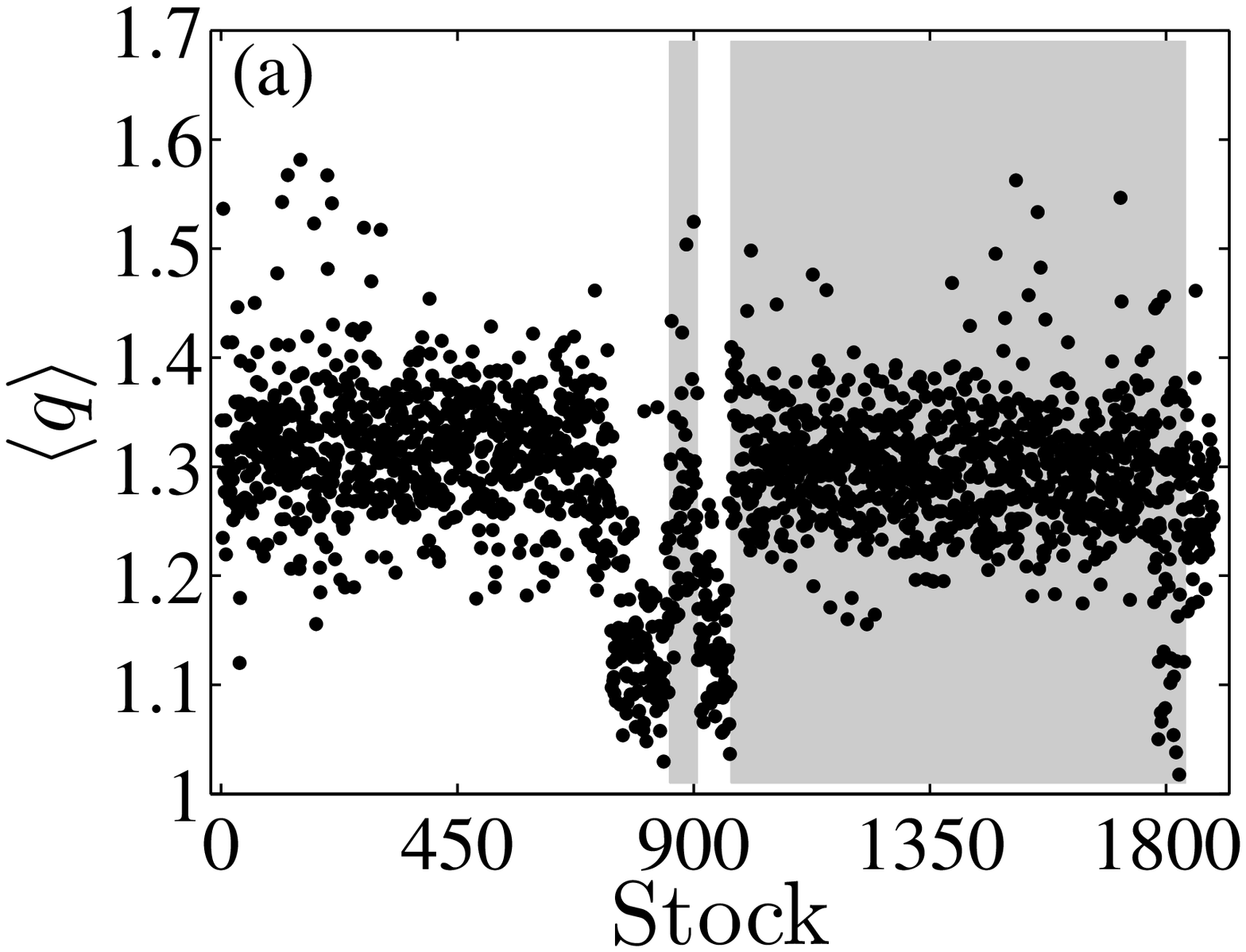}
\includegraphics[width=5cm]{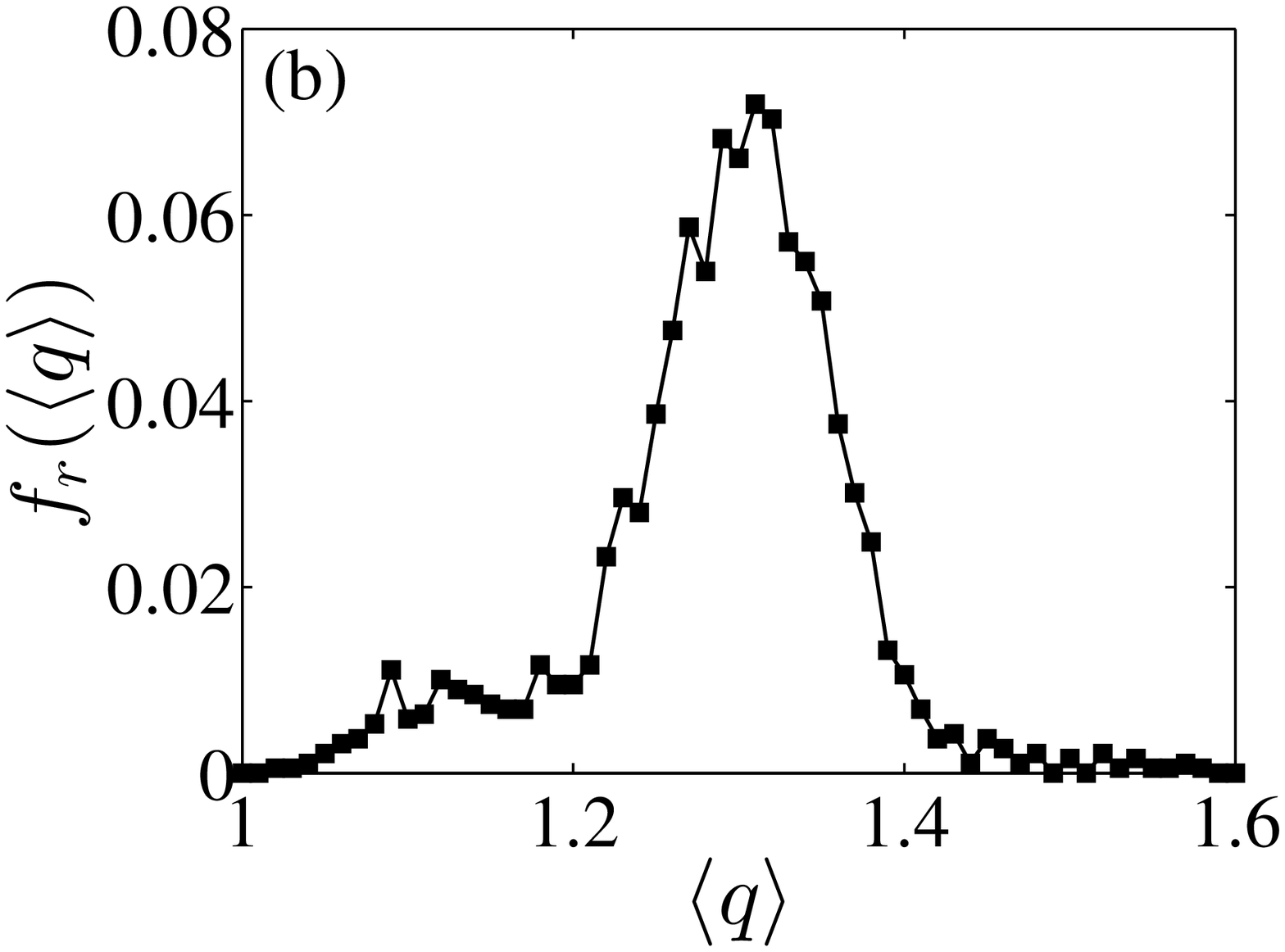}
\includegraphics[width=5cm]{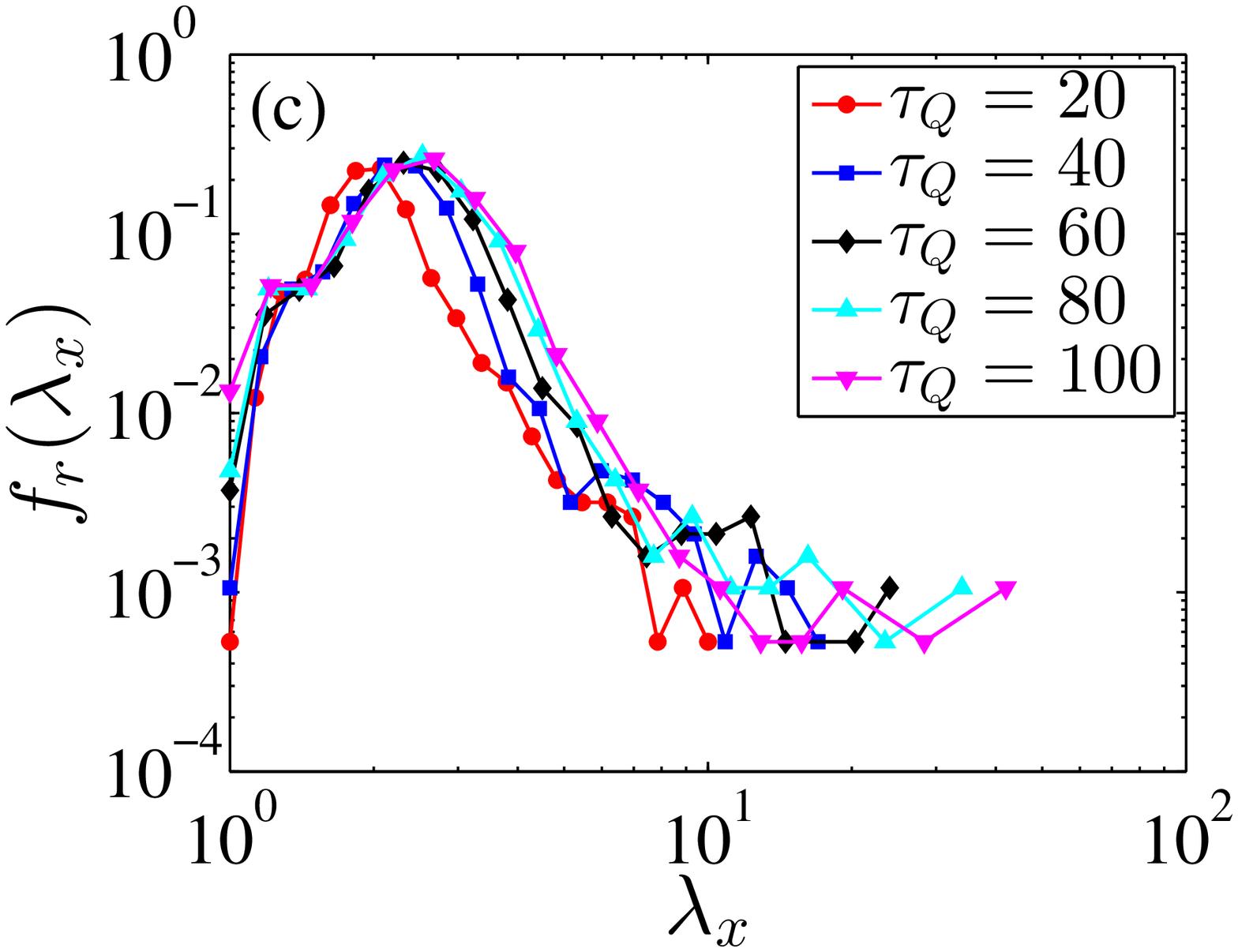}
\caption{\label{Fig:RI:Freq:FitPar} (color online). Plots of the
  estimated $q$ and $\lambda_x$ of $q$-exponential distribution for all
  stocks. (a) Plots of the fitting parameter $q$ for different
  stocks. There are five areas, three white areas are separated by two
  dark areas. For left to right, the first area represents the A-shares
  in Shenzhen market, the second area is the B-shares in Shenzhen
  market, the third area stands for ChiNext shares in Shenzhen market,
  the fourth area corresponds to the A-shares in Shanghai market, and
  the fifth area is the B-shares in Shanghai market. (b) Frequency of
  the fitting parameter $q$. (c) Frequency of the fitting parameter
  $\lambda_x$ for different values of $\tau_Q$.}
\end{figure}

Since the value of $q$ does not depend on $\tau_Q$, we average the
estimated $q$ for different $\tau_Q$ values for each stock and plot the
mean value of $q$ in Fig.~\ref{Fig:RI:Freq:FitPar}(a). Note the three
white areas separated by two shadow areas. The five areas from left to
right represent A-shares in the Shenzhen market, B-shares in the
Shenzhen market, ChiNext shares in the Shenzhen market, A-shares in the
Shanghai market, and B-shares in the Shanghai market. Note that there
are two groups of stocks, one with a relative smaller value of $\langle
q \rangle$ and the other with a much larger value of $\langle q
\rangle$. The group of stocks with the smaller $\langle q \rangle$ have
been traded in the market less than three years. The panel in
Fig.~\ref{Fig:RI:Freq:FitPar}(b) shows a plot of the frequency of
$\langle q \rangle$ in which there is a significant peak at $\langle q
\rangle\approx 1.3$, which corresponds to the average value of $q$ for
the stocks in the large $q$ group. Note also that there is a small peak
at $q\approx 1.1$, which is the mean value of $q$ for the stocks in the
small $q$ group. Because the fitting parameter $\lambda_x$ of half of
the stocks in our sample depends on $\tau_Q$, we plot the frequency of
$\lambda_x$ of all stocks for different values of $\tau_Q$ [see
  Fig.~\ref{Fig:RI:Freq:FitPar}(c)], instead of averaging the
$\lambda_x$ of different $\tau_Q$ for each stock. Note that the
distribution curve of $\lambda_x$ displays the same pattern across
different values of $\tau_Q$, the only difference being that the
distribution spanning range becomes wider when $\tau_Q$ increases. This
further indicates that the fitting parameter $\lambda_x$ is affected by
the mean recurrence time $\tau_Q$.

In order to determine whether the $q$-exponential parameters are
influenced by market state, i.e., bull or bear, we use a moving window
analysis to track the evolution of fitting parameters $q$ and
$\lambda_x$. Because stocks with trading records shorter than three
years have smaller $q$ values, we fix the window size at 48 months and
exclude stocks with trading periods shorter than 89 months.  We also
discard first-month trading data from the stocks remaining because
first-month records tend to be partial (i.e., do not span an entire
month) and the volatilities for new IPOs excessively large. We arrive at
1395 stocks as subject for our rolling window analysis.

\begin{figure}[htb]
\centering
\includegraphics[width=8cm]{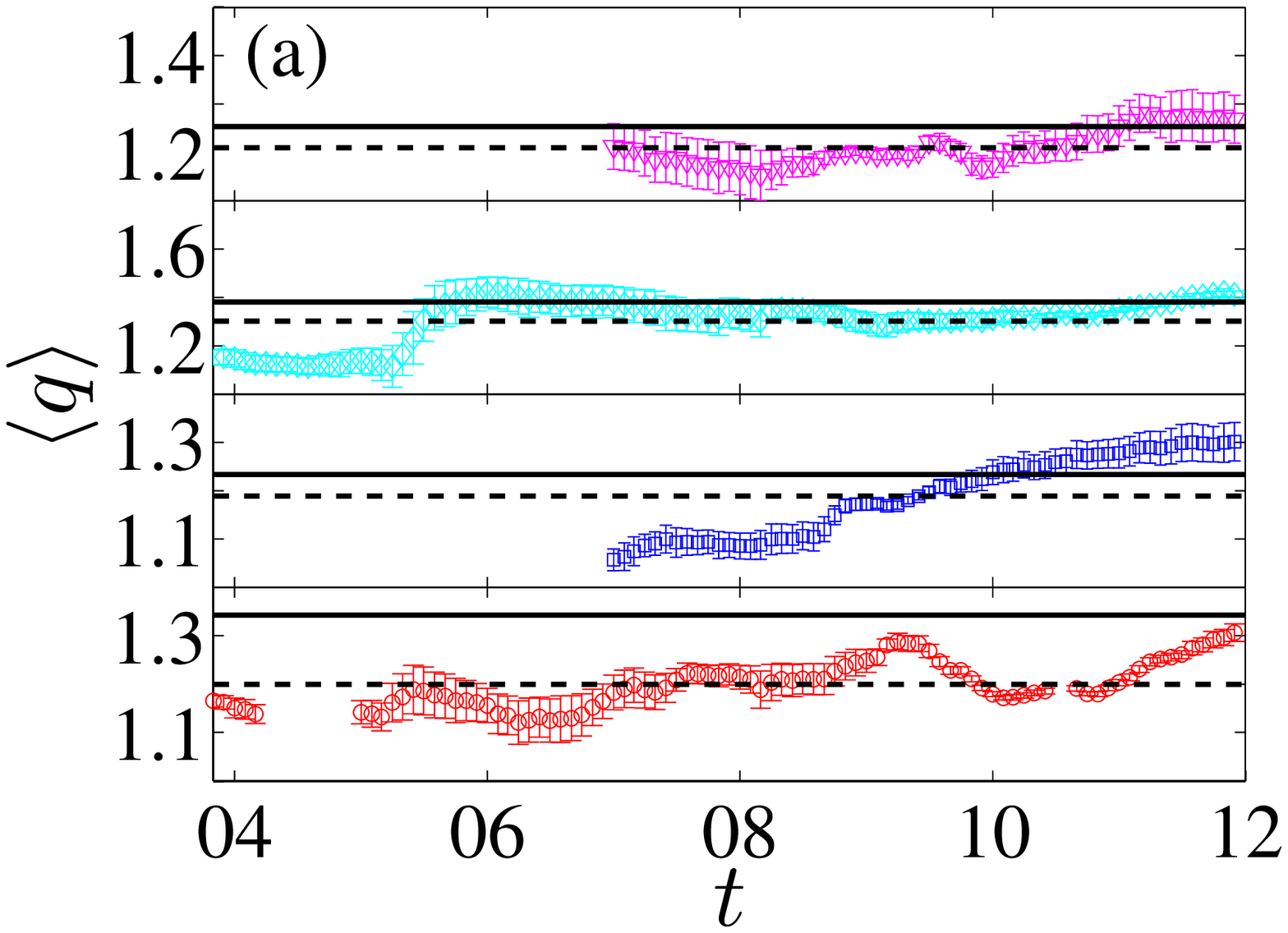}
\includegraphics[width=8cm]{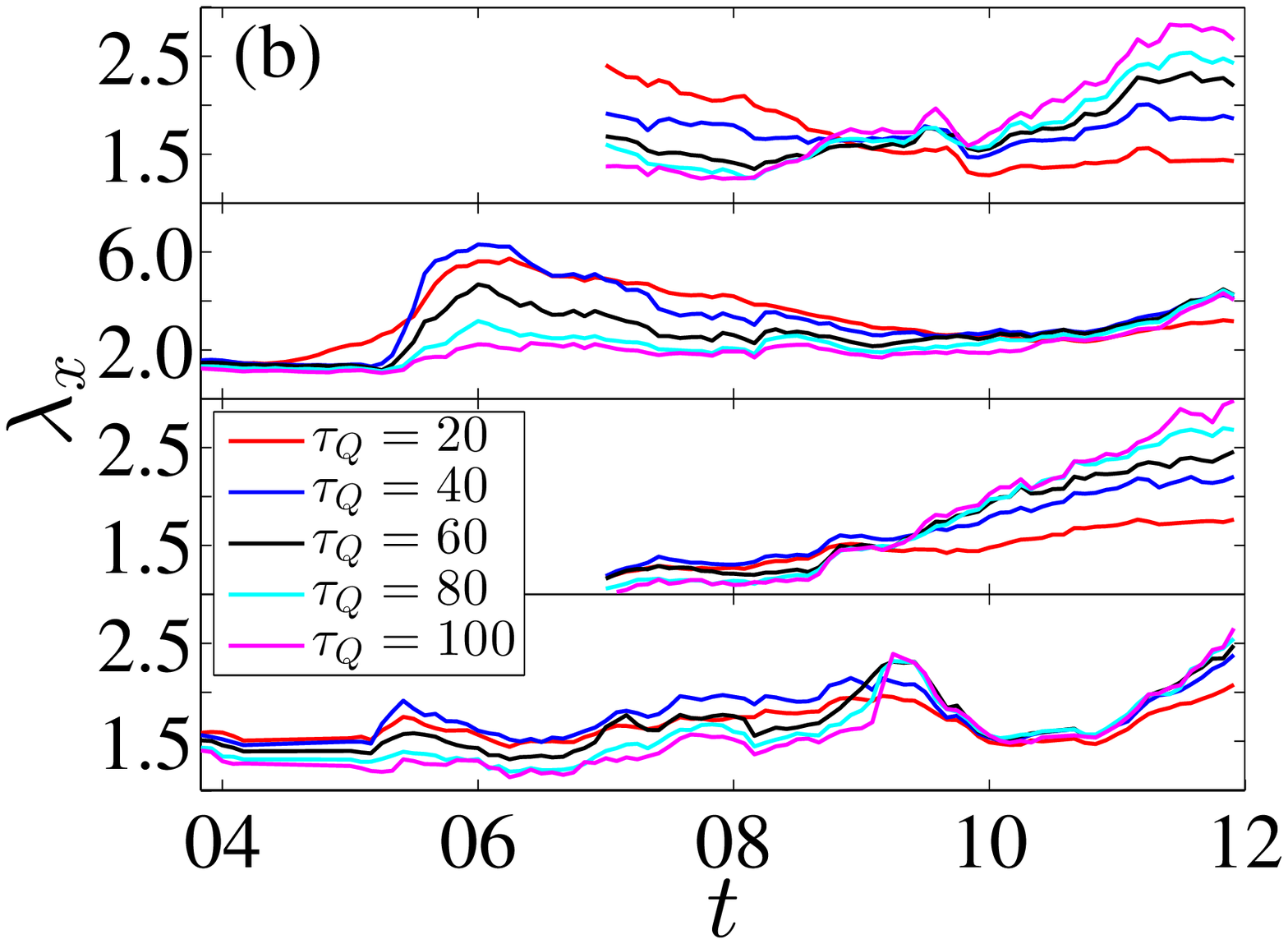}
\caption{\label{Fig:RI:Freq:FitPar:MW} (color online). Plots of the
  evolution of the estimated $q$ and $\lambda_x$ of $q$-exponential
  distribution for four chosen stocks. From the bottom panel to the up
  panel, the stock codes are 000001, 200002, 600220, and 900956,
  respectively. (a) Evolution of $q$. The dash line represents the mean
  value of $q$ across all windows. The solid line corresponds to the $q$
  of the entire period. (b) Evolution of $\lambda_x$ for different
  values of $\tau_Q$.}
\end{figure}

For each stock, we perform the same analysis in each moving window of
the series that we perform on the whole series. We first fit the
recurrence intervals for different values of $\tau_Q$ with the
$q$-exponential distribution and estimate the corresponding distribution
parameters $q$ and $\lambda_x$ in each window. We again find that the
parameter $q$ is independent of $\tau_Q$ in each window for each stock,
but the parameter $\lambda_x$ does not exhibit this behavior. We average
the $q$ for different values of $\tau_Q$ in each window and plot the
mean $q$ as a function of the time, which corresponds to the last month
of each moving window [see Fig.~\ref{Fig:RI:Freq:FitPar:MW}(a)]. For the
sake of comparison, the estimated $q$ of the entire period is a solid
line and the mean value of $q$ across all the windows is a dashed
line. Note that for the four stocks there is a big gap between the solid
and dashed line. Note also that the $\langle q \rangle$ curves also
exhibit significant fluctuations along the time
axis. Figure~\ref{Fig:RI:Freq:FitPar:MW}(b) shows the evolution of
$\lambda_x$ for different values of $\tau_Q$. Note that the trajectory
of $\lambda_x$ exhibits the same trend as the trajectory of $\langle q
\rangle$. It is not clear whether these trends reflect stock price
tends, but our results demonstrate that the estimated parameters of the
$q$-distribution are influenced by market status and thus may be treated
as a individual risk factor when explaining market returns.

\section{Results of the large volatility prediction}
\label{S1:Risk}

\subsection{Hazard probability}

The recurrence intervals between volatilities exceeding a threshold $Q$
are well approximated by the $q$-exponential distribution, and that
gives us the formula of interval distribution $p(t) = (2-q) \lambda
[1+(q-1) \lambda t]^{-\frac{1}{q-1}}$. By substituting this equation
into equation~(\ref{Eq:Wq}), we obtain
\begin{equation}
 W_q(\Delta{t}|t)= 1- \left[ 1 + \frac{(q-1)\lambda \Delta
     t}{1+(q-1)\lambda t} \right]^{1-\frac{1}{q-1}}.
 \label{Eq:Wq:qe}
\end{equation}

\begin{figure}[htb]
\centering
\includegraphics[width=5cm]{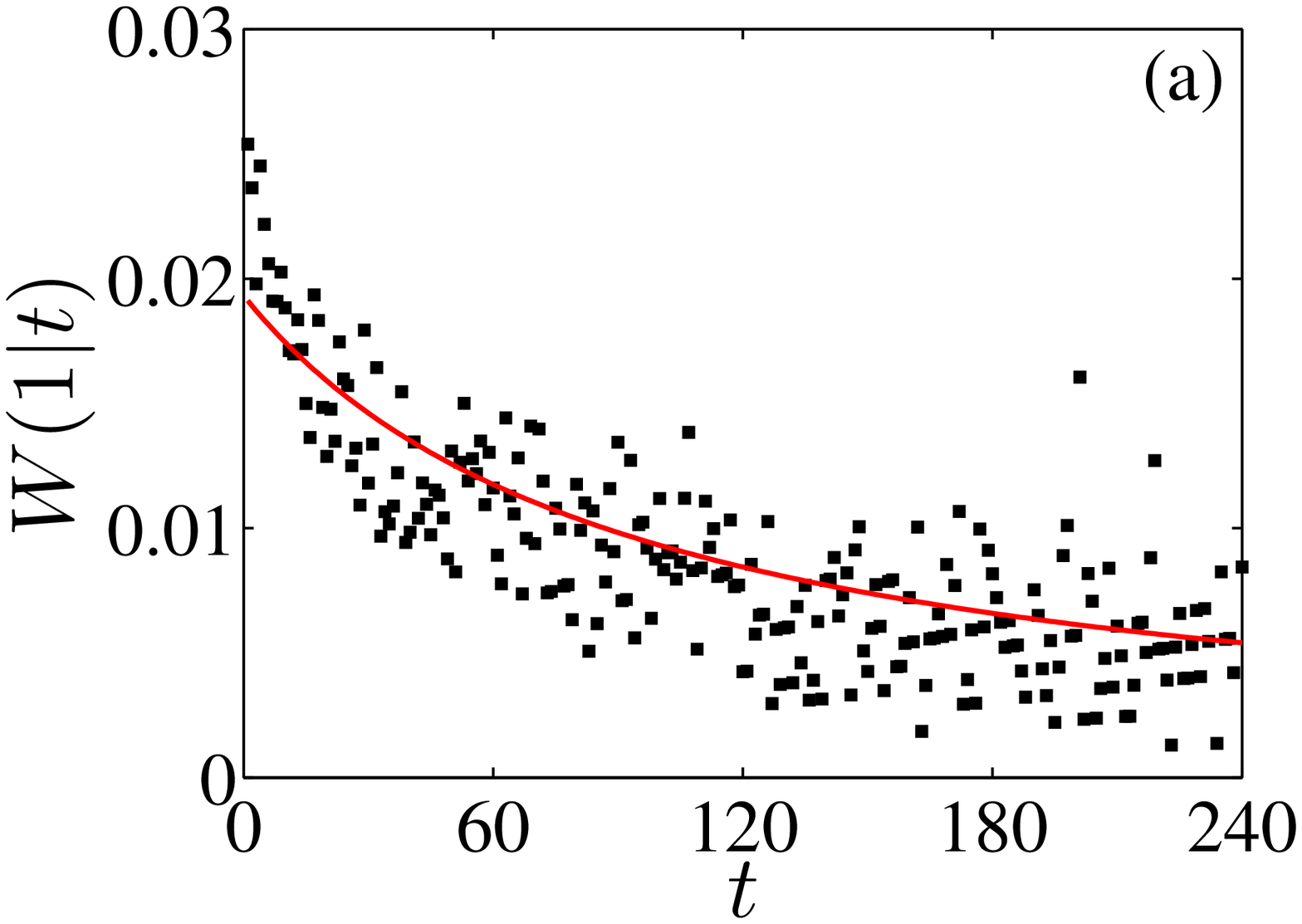}
\includegraphics[width=5cm]{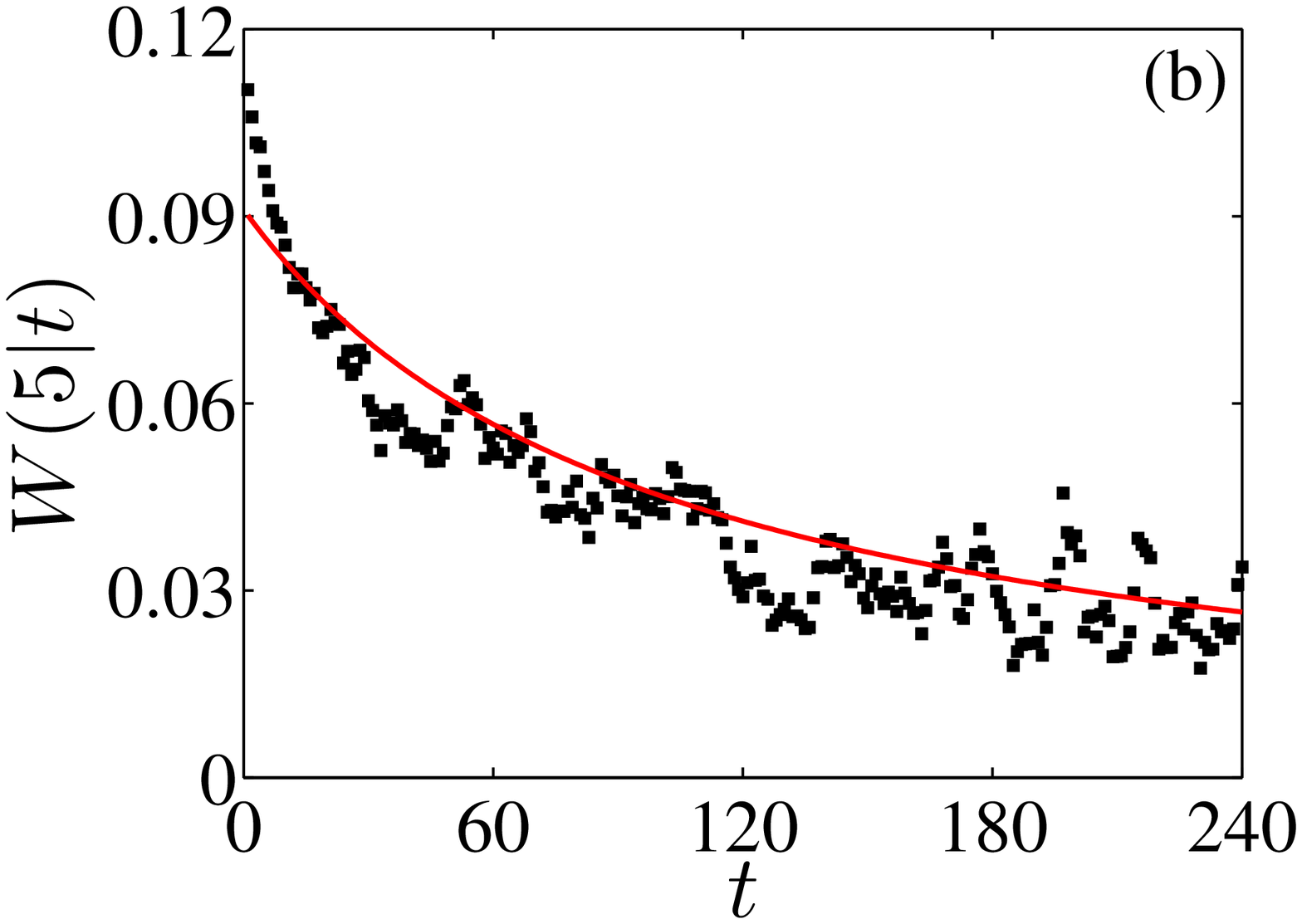}
\includegraphics[width=5cm]{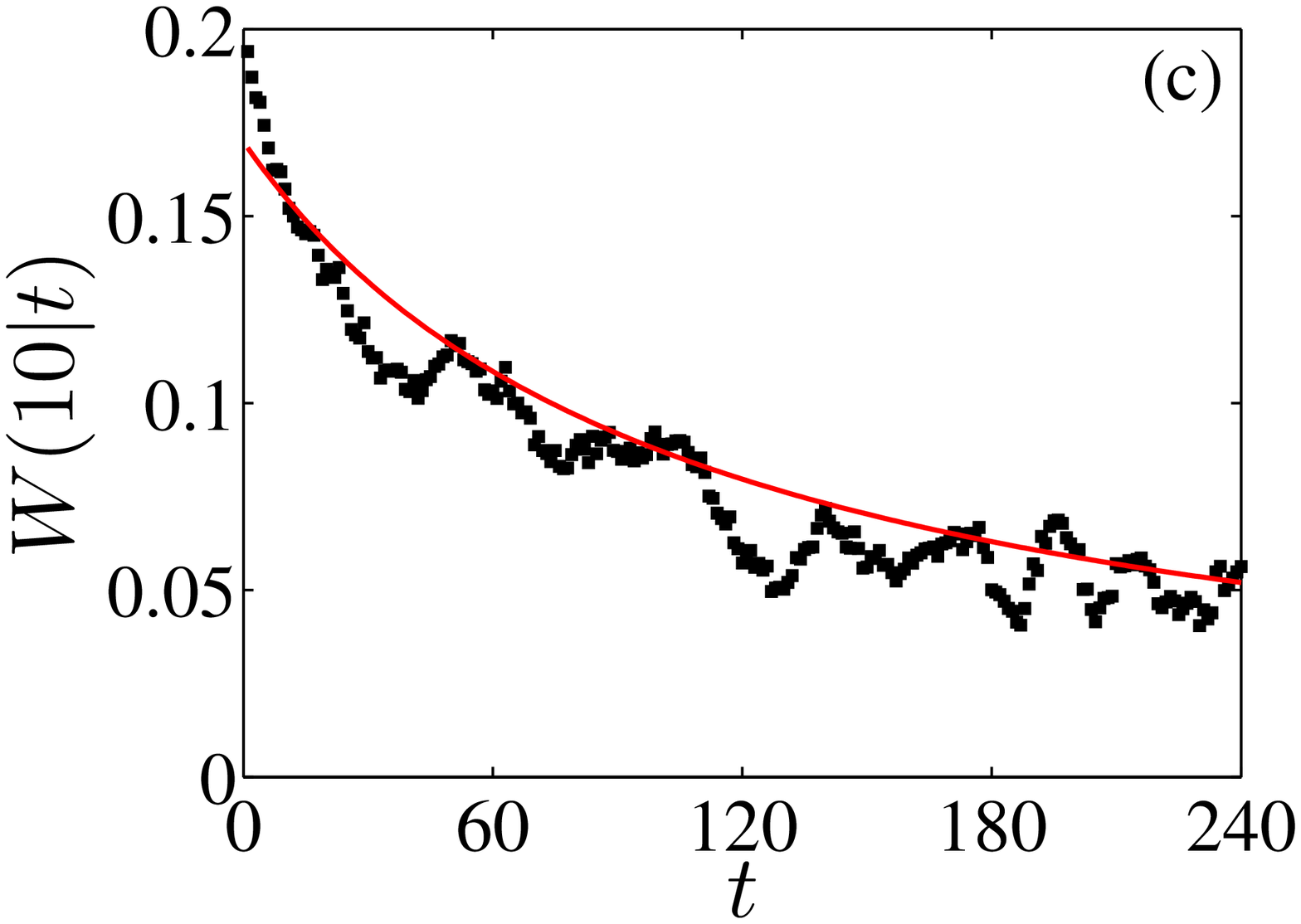}
\includegraphics[width=5cm]{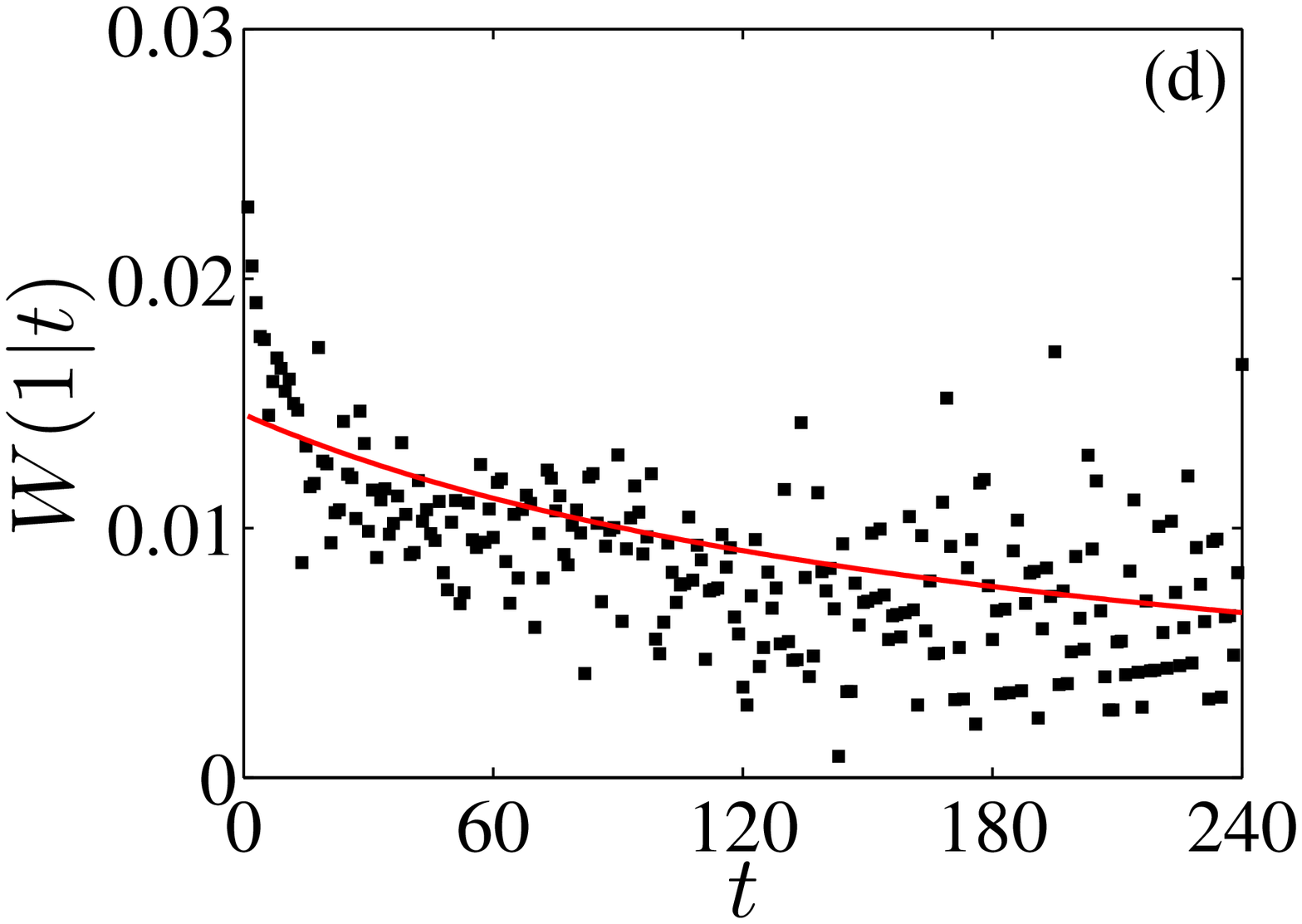}
\includegraphics[width=5cm]{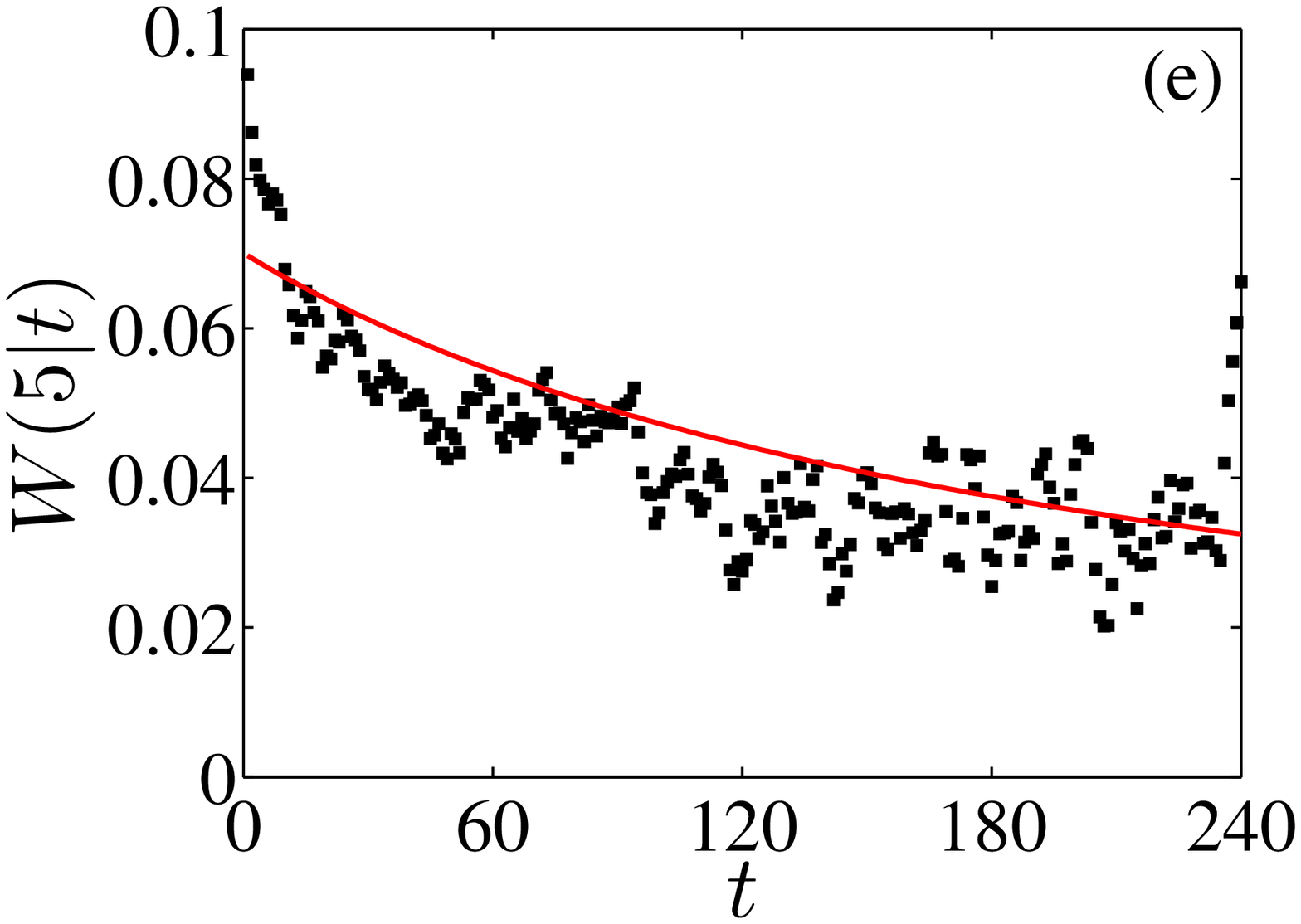}
\includegraphics[width=5cm]{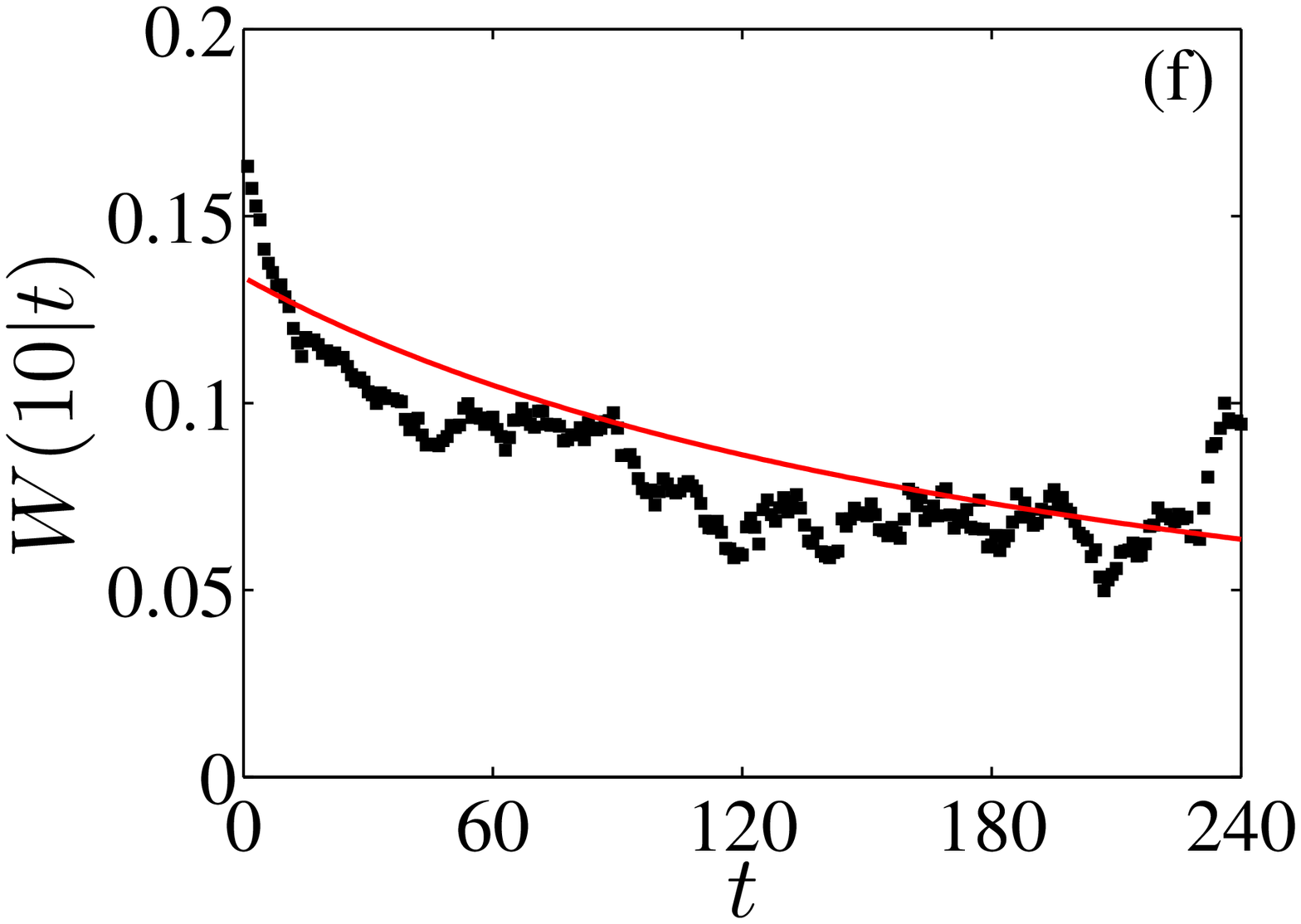}
\caption{\label{Fig:RI:Wdtt} (color online). Plot of hazard probability
  $W(\Delta t|t)$ for two stocks and $\Delta t = 1$, $5$, and
  $10$. (a-c) Stock 000001. (d-f) Stock 900956.}
\end{figure}

If we designate the top 1\% of volatility values (corresponding to the
mean recurrence time $\tau_Q = 100$) to be extreme events, we can
estimate the hazard probability $W_q(\Delta{t},t)$ in $t$ when fixing
$\Delta t$. Figure~\ref{Fig:RI:Wdtt} shows the estimated hazard
probability for two stocks when $\Delta t = 1$, 5, and 10. Note that
both the solid lines indicating the analytical solution
equation~(\ref{Eq:Wq:qe}) and the markers indicating the empirical data
decrease slowly and in each panel are in good agreement. The decreasing
trend of $W(\Delta t |t)$ is consistent with the clustering behavior in
the volatility series. Note that the hazard probability $W(\Delta t |t)$
is universal and can be used to estimate the risk in any kind of time
series.

\subsection{Predicting large volatilities}

\begin{figure}[htb]
\centering
\includegraphics[width=5.5cm]{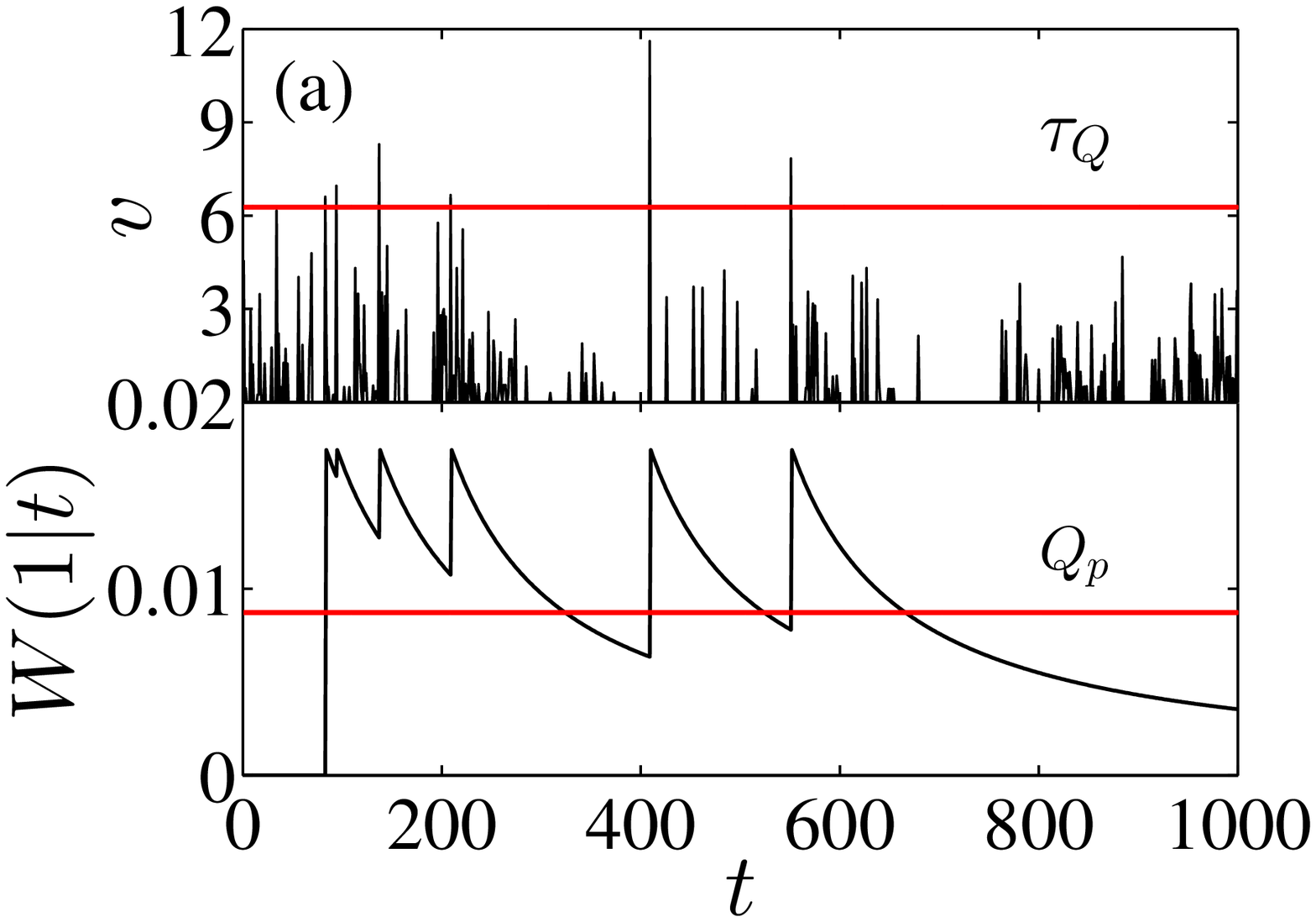}
\includegraphics[width=5cm]{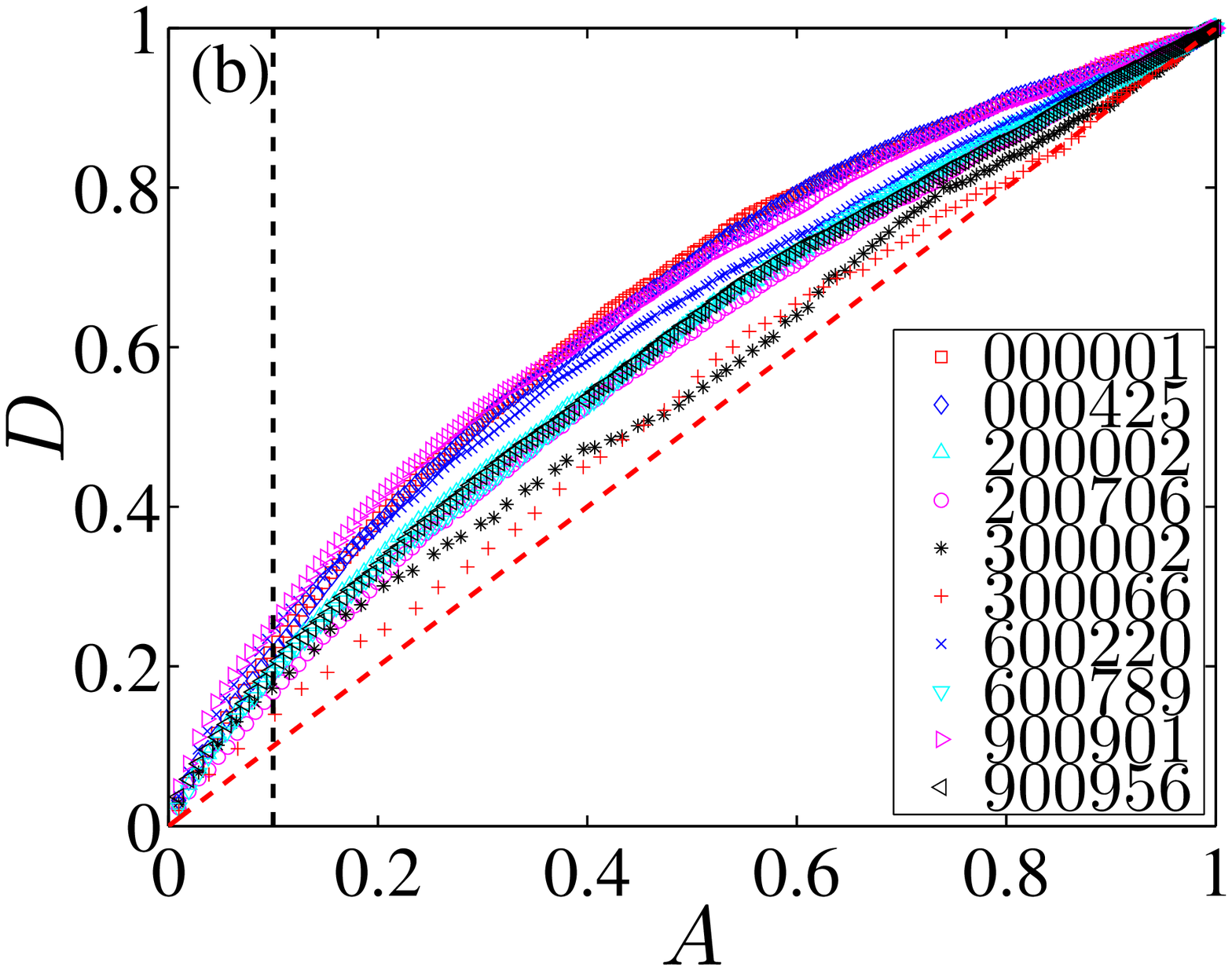}
\includegraphics[width=5cm]{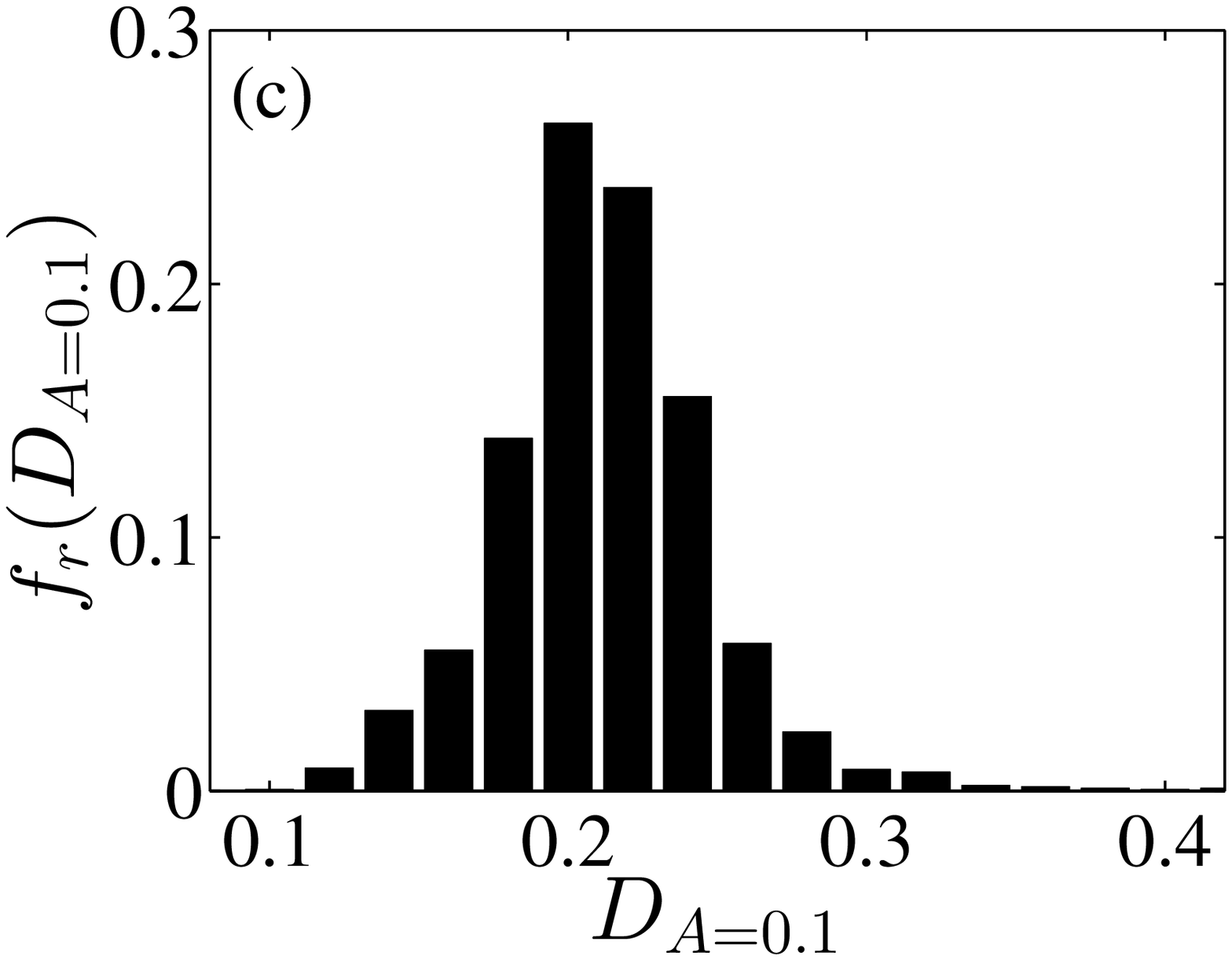}
\caption{\label{Fig:RI:PreExtE} (color online). Prediction of large
  volatilities. (a) Plots of a representative volatility series in the
  top panels and hazard probability $W(1|\Delta t)$ in the bottom
  panels. (b) Plots of ROC curves for ten stocks. (c) Distribution plots
  of the correct prediction rates at the false alarm level of 0.1 for
  all stocks.}
\end{figure}

To forecase large volatility events in a volatililty series we first
calculate the recurrence time for a given $\tau_Q$, e.g., $\tau_Q =
100$, and estimate the distribution of parameters. The hazard
probability $W(1|t)$ that an extreme event will occur in the next period
is determined using equation~(\ref{Eq:Wq:qe}) and the estimated distribution
parameters. Once the hazard probability breaks though the predefined
threshold $Q_p$, an alarm will be triggered, warning that a large
volatility event is immanent. Figure~\ref{Fig:RI:PreExtE}(a) plots a
subseries of the volatility values and highlights events above threshold
$Q$ in the top panel, which correspond to the mean recurrence time
$\tau_Q$. The risk probability $W(1|t)$ is shown in the bottom
panel. Note that $W(1|t)$ decreases as time $t$ elapsed from the last
large volatility event increases.  Threshold $Q_p$ is plotted as a
horizontal line to show the activating alarm process. By varying $Q_p$
within a $[0, 1]$ range, we obtain all pairs of $(A, D)$.

Figure~\ref{Fig:RI:PreExtE}(b) shows the ROC curves for ten stocks. Note
that all ten curves are above the dashed line $D=A$, indicating that our
prediction is not random.  The ten curves do not overlap, indicating
that the accuracy of this prediction algorithm varies for different
stocks. For the same false alarm $A$ (where a vertical dashed line is
plotted at $A=0.1$ as an example), stock 900901 has the highest correct
prediction rate and stock 300066 the lowest. We also calculate the
correct prediction rate $D_{A=0.1}$ for all stocks at the false alarm
level of 0.1. Figure~\ref{Fig:RI:PreExtE}(c) shows the frequency plots
of $D_{A={\rm{0.1}}}$. Note that the peak is centered at $\approx
0.2$, 10\% higher than a random prediction. We also find there are
nine stocks with $D_{A=0.1} > 0.4$. The top three values are 0.7, 0.68,
and 0.60 for stocks 000529, 000557, and 000592, indicating that our
forecasting algorithm can accurately predict the large volatilities of
these three stocks. Previous research has indicated that the efficiency
of the algorithm is primarily influenced by the linear and nonlinear
memory behavior in the original volatilities
\citep{Bogachev-Bunde-2011-PA}. Our results here indicate that the
behavior of stocks with stronger memory behaviors, such as volatility
clustering and multifactality, could be more accurately predicted using
our algorithm. Our algorithm only takes into consideration the
probability distribution of recurrence intervals, but if the memory
behavior of recurrence intervals were also included we believe that its
predictive accuracy would be greatly inhanced.

\section{Conclusion}
\label{S1:Conclusion}

In this work, we have utilized a decision-making algorithm to forecast
the occurrence of large volatilities in Chinese stock markets based on
the hazard probability, which is derived from the distribution of
recurrence intervals between the volatilities exceeding a threshold
$Q$. By fitting the volatility recurrence intervals by means of five
candidate distributions and comparing their KS statistics, we have found
that the volatility recurrence intervals are well approximated by the
$q$-exponential distribution. The fitting parameter $q$ is found to be
independent of the mean recurrence time $\tau_q$, which is at a
one-to-one correspondence with the threshold $Q$, for all the stocks in
our sample. However the parameter $\lambda_x$ of the $q$-distribution
does not exhibit the same behavior as $q$. For half of the stocks,
$\lambda_x$ exhibits a strong dependence on $\tau_Q$. Using a moving
window analysis, we have found that both parameters are influenced by
market status and exhibit the same trend with the evolution of
time. This behavior may have potential applications for explaining stock
return volitility.

Using the $q$-distribution formula, we have derived an analytical
solution of the hazard probability $W(\Delta t|t)$ of the next large
volatility event above the threshold $Q$ within a short time interval
$\Delta t$ after an elapsed time $t$ from the last large volatility
above $Q$. This analytical solution $W(\Delta t|t)$ is in good agreement
with the empirical risk probability derived from real stock data. We
have adopted a decision-marking algorithm and have used the hazard
probability to forecast large volatilities. At the false alarm level of
0.1, we have found that the average correct predicting rate is 0.2 for
all stocks. We have also found that there are three stocks with a
correct predicting rate is greater than 0.6, indicating that our
predicting algorithm is accurate in forecasting the large
volatilities. Our findings may shed new light on our understanding of
extreme volatility behavior and may have potential applications in
managing stock market risk.

\section*{Acknowledgements}

Z.-Q.J. and W.-X.Z. acknowledge support from the National Natural
Science Foundation of China (71131007), Shanghai ``Chen Guang'' Project
(2012CG34), Program for Changjiang Scholars and Innovative Research Team
in University (IRT1028), China Scholarship Council (201406745014) and
the Fundamental Research Funds for the Central
Universities. A.C. acknowledges the support from Brazilian agencies
FAPEAL (PPP 20110902-011-0025-0069/60030-733/2011) and CNPq (PDE
20736012014-6).

\bibliographystyle{elsarticle-num}
\bibliography{E:/Papers/Auxiliary/Bibliography}

\appendix

\section{Maximum likelihood estimation of distribution parameters}

\subsection{Stretched exponential distribution}

To estimate the parameters of Eq.~(\ref{Eq:StrExp:x:PDF}), the first step is to use $\int_0^\infty f(x) = 1$ and $\int_0^\infty xf(x) = 1$ to reduce the number of parameters.For the first integral, we have
\begin{equation}
\int_0^\infty f(x) {\rm{d}}{x} = \int_0^\infty a \tau_Q e^{-(b \tau_Q x)^\mu} {\rm{d}}{x} = 1,
 \label{Eq:StrExp:Fit1}
\end{equation}
Let $y = (b\tau_Q x)^\mu$, we have $x = \frac{y^{1/\mu}}{b\tau_Q}$ and ${\rm{d}} x = \frac{ y^{1/\mu-1} {\rm{d}} y}{\mu b\tau_Q }$. Then, we can obtain
\begin{equation}
\int_0^\infty a \tau_Q e^{-(b \tau_Q x)^\mu} {\rm{d}}{x} = \int_0^\infty a\tau_Q e^{-y} \frac{ y^{1/\mu-1}}{\mu b\tau_Q } {\rm{d}} y
= \frac{a}{\mu b} \int_0^\infty y^{1/\mu-1} e^{-y} {\rm{d}} y = \frac{a}{\mu b} \Gamma(\frac{1}{\mu})=1.
 \label{Eq:StrExp:Fit2}
\end{equation}
For the second integral, we have
\begin{equation}
\int_0^\infty xf(x) {\rm{d}}{x} = \int_0^\infty x a \tau_Q e^{-(b \tau_Q x)^\mu} {\rm{d}}{x} = 1,
 \label{Eq:StrExp:Fit3}
\end{equation}
Again, we let $y = (b\tau_Q x)^\mu$, we have $x = \frac{y^{1/\mu}}{b\tau_Q}$ and ${\rm{d}} x = \frac{ y^{1/\mu-1} {\rm{d}} y}{\mu b\tau_Q }$. Then, we can obtain
\begin{equation}
\int_0^\infty a x \tau_Q e^{-(b \tau_Q x)^\mu} {\rm{d}}{x} = \int_0^\infty \frac{y^{1/\mu}}{b\tau_Q} a\tau_Q e^{-y} \frac{ y^{1/\mu-1}}{\mu b\tau_Q } {\rm{d}} y
= \frac{a}{\mu b^2 \tau_Q} \int_0^\infty y^{2/\mu-1} e^{-y} {\rm{d}} y = \frac{a}{\mu b^2 \tau_Q} \Gamma(\frac{2}{\mu})=1.
 \label{Eq:StrExp:Fit4}
\end{equation}
Through solving the equations, the parameters $a$ and $b$ could be formulated by the exponent $\mu$ and $\tau_Q$,
\begin{equation}
a = \frac{\mu \Gamma(2/\mu)}{\Gamma(1/\mu)^2 \tau_Q}, ~~ b = \frac{\Gamma(2/\mu)}{\Gamma(1/\mu) \tau_Q}
 \label{Eq:StrExp:Fit5}
\end{equation}

The likelihood function of the stretched exponential distribution can be written as
\begin{equation}
L = \prod_i^n a \tau_Q \exp[-(b \tau_Q x_i)^{\mu}],
 \label{Eq:StrExp:Likelihood}
\end{equation}
Taking logarithm on both side, we have
\begin{equation}
\ln L =  n \ln a + n \ln \tau_Q  - \sum_i^n (b \tau_Q x_i)^{\mu},
 \label{Eq:StrExp:Likelihood}
\end{equation}
By submitting Eq.~(\ref{Eq:StrExp:Fit5}) into Eq.~(\ref{Eq:StrExp:Likelihood}), the log likelihood function of the stretched exponential distribution has only one variable $\mu$. Our purpose is to find the value of $\mu$ which is associated with the maximum value of the $\ln L$. Here, it is very hard to obtain the expression by taking a derivative of $\ln L$ with respect to $\mu$. Hence, we just estimate the function value of $\ln L$ by changing $\mu$ from 0 to 5 with a step of $10^{-6}$. We locate the $\mu$ with the maximum $\ln L$ as the solution of our maximum likelihood estimation.

\subsection{Power-law distribution with an exponential cutoff}

As the same way as the stretched exponential, we use $\int_0^\infty f(x) = 1$ and $\int_0^\infty xf(x) = 1$ to reduce the number of parameters. For the first integral, we have
\begin{equation}
\int_0^\infty f(x) {\rm{d}} x = \int_0^\infty c \tau_Q^{-\gamma} x^{-\gamma-1} e^{-k \tau_Q x} {\rm{d}} x = 1,
 \label{Eq:PowExp:Fit1}
\end{equation}
Let $y = k \tau_Q x$, we have $x = \frac{y}{k\tau_Q}$ and ${\rm{d}} x = \frac{{\rm{d}}y}{k\tau_Q}$. Then, we obtain
\begin{equation}
\int_0^\infty c \tau_Q^{-\gamma} x^{-\gamma-1} e^{-k \tau_Q x} {\rm{d}} x = \int_0^\infty c \tau_Q^{-\gamma} \left(\frac{y}{k\tau_Q}\right) ^{-\gamma-1}  e^{-y} \frac{{\rm{d}}y}{k\tau_Q} = \frac{c}{k^{-\gamma}} \int_0^\infty y^{-\gamma-1} e^{-y} =  \frac{c}{k^{-\gamma}} \Gamma(-\gamma) = 1.
 \label{Eq:PowExp:Fit2}
\end{equation}
For the second integral, we have
\begin{equation}
\int_0^\infty xf(x) {\rm{d}} x = \int_0^\infty x c \tau_Q^{-\gamma} x^{-\gamma-1} e^{-k \tau_Q x} {\rm{d}} x = 1,
 \label{Eq:PowExp:Fit1}
\end{equation}
Again, we let $y = k \tau_Q x$, we have $x = \frac{y}{k\tau_Q}$ and ${\rm{d}} x = \frac{{\rm{d}}y}{k\tau_Q}$. Then, we obtain
\begin{equation}
\int_0^\infty x c \tau_Q^{-\gamma} x^{-\gamma-1} e^{-k \tau_Q x} {\rm{d}} x = \int_0^\infty c \frac{y}{k\tau_Q} \tau_Q^{-\gamma} \left(\frac{y}{k\tau_Q}\right) ^{-\gamma-1}  e^{-y} \frac{{\rm{d}}y}{k\tau_Q} = \frac{c}{k^{1-\gamma} \tau_Q} \int_0^\infty y^{(1-\gamma)-1} e^{-y} =  \frac{c}{k^{1-\gamma } \tau_Q} \Gamma(1-\gamma) = 1.
 \label{Eq:PowExp:Fit4}
\end{equation}
Through solving the two equations, we obtain that
\begin{equation}
k = \frac{\Gamma(1-\gamma)}{\Gamma(-\gamma) \tau_Q} = \frac{-\gamma}{\tau_Q}, ~~ b = \left(\frac{-\gamma}{\tau_Q}\right)^{-\gamma} \frac{1}{\Gamma(-\gamma)}
 \label{Eq:PowExp:Fit5}
\end{equation}

The likelihood function of the power-law distribution with an exponential cutoff can be written as
\begin{equation}
L = \prod_i^n c \tau_Q^{-\gamma} x^{-\gamma-1} e^{-k \tau_Q x},
 \label{Eq:PowExp:Likelihood}
\end{equation}
Taking logarithm on both side, we have
\begin{equation}
\ln L =  n \ln c - \gamma n \ln \tau_Q  - (\gamma+1) \sum_i^n \ln x_i- \sum_i^n k \tau_Q x_i,
 \label{Eq:SoeExp:Likelihood}
\end{equation}
By submitting Eq.~(\ref{Eq:PowExp:Fit5}) into Eq.~(\ref{Eq:SoeExp:Likelihood}), the log likelihood function of the power-law distribution with an exponential distribution has only one variable $\gamma$. Our purpose is to find the value of $\gamma$ which is associated with the maximum value of the $\ln L$. Here, it is very hard to obtain the expression by taking a derivative of $\ln L$ with respect to $\gamma$. Hence, we just estimate the function value of $\ln L$ by changing $\mu$ from -1 to 0 with a step of $10^{-6}$. We locate the $\gamma$ with the maximum $\ln L$ as the solution of our maximum likelihood estimation.







\end{document}